\documentclass[useAMS,usenatbib]{mn2e}
\usepackage{graphicx}
\DeclareMathAlphabet{\mathpzc}{OT1}{pzc}{m}{it}

\def\lsim{\,\lower2truept\hbox{${<\atop\hbox{\raise4truept\hbox{$\sim$}}}$}\,}
\def\gsim{\,\lower2truept\hbox{${> \atop\hbox{\raise4truept\hbox{$\sim$}}}$}\,}

\title[ {\it Planck} local luminosity function]{The local luminosity
  function of star-forming galaxies derived from the {\it Planck} Early Release Compact Source Catalogue}
\author[M. Negrello et al.]
{M.~Negrello$^{1}$\thanks{mattia.negrello@oapd.inaf.it.},
M.~Clemens$^{1}$,
J.~Gonzalez-Nuevo$^{2}$,
G.~De Zotti$^{1,3}$,
L.~Bonavera$^{2}$,
\newauthor
G.~Cosco$^{4}$,
G.~Guarese$^{4}$,
L.~Boaretto$^{4}$,
S.~Serjeant$^{5}$,
L.~Toffolatti$^{6}$,
A.~Lapi$^{3,7}$,
\newauthor
M.~Bethermin$^{8}$,
G.~Castex$^{9}$,
D.~L.~Clements$^{10}$,
J.~Delabrouille$^{9}$,
H.~Dole$^{11,12}$,
\newauthor
A.~Franceschini$^{13}$,
N.~Mandolesi$^{14,15}$,
L.~Marchetti$^{13}$,
B.~Partridge$^{16}$,
A. Sajina$^{17}$ \\
$^{1}$INAF, Osservatorio Astronomico di Padova, Vicolo Osservatorio 5,
I-35122 Padova, Italy \\
$^{2}$ Inst. de F{\'\i}sica de Cantabria (CSIC-UC), Avda. los Castros s/n,
39005 Santander, Spain \\
$^{3}$SISSA, Via Bonomea 265, I-34136 Trieste, Italy \\
$^{4}$Gruppo Astrofili Polesani, Osservatorio Astronomico Vanni Bazzan di Sant'Apollinare, I-45100 Rovigo, Italy \\
$^{5}$Department of Physical Sciences, The Open University, Milton Keynes MK7 6AA, United Kingdom \\
$^{6}$Departamento de Fisica, Universidad de Oviedo, Avda. Calvo Sotelo s/n, 33007 Oviedo, Spain \\
$^{7}$Dipartimento di Fisica, Universit\`a `Tor Vergata', Via Ricerca Scientifica 1, 00133 Roma, Italy \\
$^{8}$Laboratoire AIM, IRFU/Service d'Astrophysique - CEA/DSM - CNRS - Universit\'{e} Paris Diderot, B$\hat{a}$t. 709, \\
CEA-Saclay, F-91191 Gif-sur-Yvette Cedex, France \\
$^{9}$APC, 10, rue Alice Domon et L\'eonie Duquet, 75205 Paris Cedex 13, France \\
$^{10}${Astrophysics Group, Imperial College, Blackett Laboratory, Prince Consort Road, London SW7 2AZ, UK} \\
$^{11}$Institut d'Astrophysique Spatiale, CNRS (UMR 8617) Universit\'{e} Paris-Sud 11, B$\hat{a}$timent 121, Orsay, France \\
$^{12}$Institut Universitaire de France, 103, bd Saint-Michel, 75005, Paris, France \\
$^{13}$Dipartimento di Fisica e Astronomia, Universit\`a di Padova, Vicolo dell'Osservatorio 3, I-35122 Padova, Italy \\
$^{14}$INAF/IASF Bologna, Via Gobetti 101, Bologna, Italy \\
$^{15}$Agenzia Spaziale Italiana, Viale Liegi 26, Roma, Italy \\
$^{16}$Haverford College, Astronomy Department, 370 Lancaster Avenue, Haverford, Pennsylvania, USA \\
$^{17}$Department of Physics and Astronomy, Tufts University, Medford, MA 02155, U.S.A. 
}
\date{Released 2012 Xxxxx XX}

\pagerange{\pageref{firstpage}--\pageref{lastpage}} \pubyear{2002}

\def\LaTeX{L\kern-.36em\raise.3ex\hbox{a}\kern-.15em
    T\kern-.1667em\lower.7ex\hbox{E}\kern-.125emX}
\def\simlt{\mathrel{\rlap{\lower 3pt\hbox{$\sim$}}\raise 2.0pt\hbox{$<$}}}
\def\simgt{\mathrel{\rlap{\lower 3pt\hbox{$\sim$}}\raise 2.0pt\hbox{$>$}}}

\begin{document}

\label{firstpage}

\maketitle

\begin{abstract}
The {\it Planck} Early Release Compact Source Catalog (ERCSC) has offered the first opportunity to accurately determine the luminosity function of dusty galaxies in the very local Universe (i.e. distances $\lsim100\,$Mpc), at several (sub-)millimetre wavelengths, using blindly selected samples of low redshift sources, unaffected by cosmological evolution. This project, however, requires careful consideration of a variety of issues including the choice of the appropriate flux density measurement, the separation of dusty galaxies from radio sources and from Galactic sources, the correction for the $CO$ emission, the effect of density inhomogeneities, and more. We present estimates of the local luminosity functions at 857\,GHz (350\,$\mu$m), 545\,GHz (550\,$\mu$m) and 353\,GHz (850\,$\mu$m) extending across the characteristic luminosity $L_\star$, and a preliminary estimate over a limited luminosity range at 217\,GHz (1382\,$\mu$m). At 850$\,\mu$m and for luminosities $L\gsim L_{\star}$ our results agree with previous estimates, derived from the SCUBA Local Universe Galaxy Survey (SLUGS), but are higher than the latter at $L\lsim L_\star$. We also find good agreement with estimates at 350 and 500\,$\mu$m based on preliminary {\it Herschel} survey data.
\end{abstract}


\section{Introduction}

Our knowledge of the (sub-)millimetre luminosity function of galaxies has substantially improved in the last few years. Before the launch of the {\it Herschel} Space Observatory (Pilbratt et al. 2010), no blind sub-mm surveys of sufficient area were available, and estimates of the local luminosity functions had to rely on follow-up of samples selected at other wavelengths. The most notable example is the local $850\,\mu$m luminosity function derived from the SCUBA Local Universe Galaxy Survey (SLUGS; Dunne et al. 2000) that provided SCUBA photometry of 104 galaxies drawn from the IRAS Bright Galaxy Survey (Soifer et al. 1989). Better constraints, particularly on the faint end of the $850\,\mu$m luminosity function as well as estimates of the luminosity function at many wavelengths in the range $60\,\mu$m--$850\,\mu$m, were obtained by Serjeant \& Harrison (2005) by modeling the spectral energy distributions (SEDs) of all 15,411 {\it IRAS} PSC$z$ galaxies (Saunders et al. 2000). The SEDs were constrained by all available far-infrared and sub-mm colour-colour relations from the SLUGS and elsewhere. The {\it Herschel} surveys have allowed the first determinations of local luminosity functions at 250, 350 and $500\,\mu$m based on complete samples of sub-mm selected galaxies (Dye et al. 2010, Vaccari et al. 2010, Dunne et al. 2011). The results are in generally good agreement with the estimates by Serjeant \& Harrison (2005).

The {\it Planck} (Planck Collaboration I, 2011) surveys have allowed the construction of the first all-sky catalogs homogeneously selected at several wavelengths from $350\,\mu$m to 1 cm, the Early Release Compact Source Catalog (ERCSC; Planck Collaboration VII, 2011). At sub-mm wavelengths the dominant population of extragalactic sources consists of nearby dusty galaxies, mostly at distances $\simlt 100\,$Mpc, all with spectroscopic redshift measurements. These are almost ideal characteristics for estimating local luminosity functions. In contrast, because of the much smaller covered areas, to get sufficient statistics with {\it Herschel} surveys it is necessary to select galaxies up to $z=0.1$--0.2 where evolutionary effects may be non-negligible, requiring model-dependent corrections. In addition, for many relatively distant galaxies only photometric redshifts are available.

On the other hand, very low redshift galaxy samples ($z<<0.1$), like those provided by {\it Planck} surveys, have their own drawbacks that must be dealt with carefully. First, proper motions can give large contributions to the measured redshifts, hence making them unreliable as distance indicators; redshift-independent distance indicators need then to be used as far as possible. Second, we live in the outskirts of the Virgo super-cluster implying that {\it Planck} samples contain strong density inhomogeneities, while the standard method to estimate the luminosity function [the $1/V_{\rm max}$ method (Schmidt 1968)] assumes a uniform distribution of galaxies. Although the effect of inhomogeneities is substantially mitigated by the (almost) full sky coverage, it must be taken into account.

In this paper we exploit the {\it Planck} ERCSC to estimate the local luminosity functions of star-forming galaxies at 217\,GHz (1382\,$\mu$m), 353\,GHz (850$\,\mu$m), 545\,GHz (550$\,\mu$m) and 857\,GHz (350$\,\mu$m).  Our analysis provides a useful complement to
the recent paper by the {\it Planck} collaboration (Planck Collaboration VII, 2012), in which the sub-millimetre to millimetre number counts and spectral indices of dust-dominated galaxies, as well as of extragalactic radio sources, selected in the ERCSC are presented.

The paper is organized as follows. In \S\,\ref{sec:sample} we describe the selection of the samples, their completeness and the correction of the {\it Planck} fluxes for CO-line emission. In \S\,\ref{sec:LF} we deal with the methodology used to measure the luminosity function and present the results, that, in \S\,\ref{sec:results}, are compared with earlier estimates. Our main conclusions are summarized in \S\,\ref{sec:conclusions}. Throughout this paper we adopt a flat cosmology  with $\Omega_{0,m}=0.27$ and $H_{0}=70\,$km\,s$^{-1}$\,Mpc$^{-1}$.

\begin{figure*}
\hspace{+0.0cm}
\makebox[\textwidth][c]{
\includegraphics[width=0.38\textwidth, angle=90]{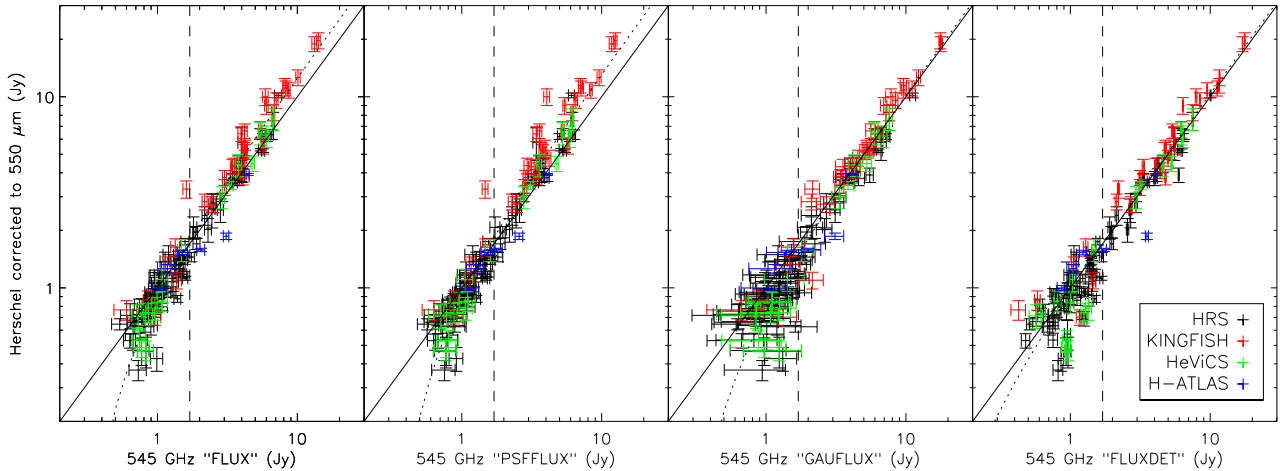}
}
\vspace{-0.5cm}
\caption{Comparison of the 4 {\it Planck} flux density estimations at 545\,GHz ($550\,\mu$m) with {\it Herschel} measurements at 500\,$\mu$m, colour corrected to 550\,$\mu$m using the spectral index determined from the {\it Planck}  measurements for each individual galaxy at 545 and 857\,GHz. The HRS flux densities (black) are from Ciesla et al. (2012), Kingfish flux densities (red) are from Dale et al. (2012), the HeViCS flux densities are from Davies et al. (2012) and the H-ATLAS flux densities (blue) are from Herranz et al. (2012). The 4 panels refer to the 4 estimates provided in the ERCSC: 'FLUX', 'PSFFLUX', 'GAUFLUX' and 'FLUXDET' (from left to right), see text.
The black solid lines correspond to a {\it Herschel}/{\it Planck} flux density ratio of 1. The dotted curves are the linear least squares fits [eq.~(\protect\ref{eq:fit})] for the full sample (coefficients $A$ and $B$ in Table~\protect\ref{tab:coefficients_rms}) and the vertical dashed lines  correspond to the adopted 80\% completeness limit (see \S\,\protect\ref{subsec:diff_counts}).
}
 \label{fig:545GHz_comp}
\end{figure*}
%

\begin{figure*}
\hspace{+0.0cm}
\makebox[\textwidth][c]{
\includegraphics[width=0.38\textwidth, angle=90]{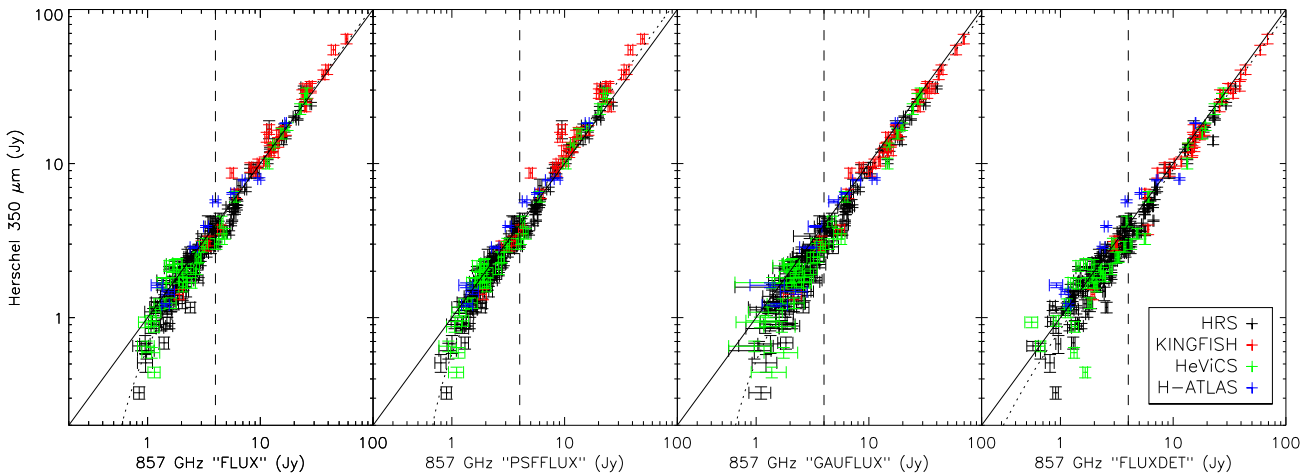}
}
\vspace{-0.5cm}
\caption{Comparison of the 4 {\it Planck} flux density estimations at 857 GHz ($350\,\mu$m) with {\it Herschel} measurements at the same frequency. The symbols and the lines have the same meaning as in Fig.~\protect\ref{fig:545GHz_comp}. }
 \label{fig:857GHz_comp}
\end{figure*}
%

\section{The samples}\label{sec:sample}

\subsection{The ERCSC}\label{sect:ERCSC}

The {\it Planck} ERCSC lists all high-reliability sources, both Galactic and extragalactic, based on mapping the entire sky once and 60\% of the sky a second time in 9 frequency bands centered at 30, 44, 70, 100, 143, 217, 353, 545 and 857 GHz. At the frequencies of interest here, the Full Width at Half Maximum (FWHM) of the {\it Planck} beam is 4.68 arcmin at 217\,GHz (1382$\,\mu$m), 4.43 arcmin at 353\,GHz (850$\,\mu$m), 3.80 arcmin at 545\,GHz (550$\,\mu$m) and 3.67 arcmin at 857\,GHz (350$\,\mu$m) (Planck Collaboration I, 2011). Below 353\,GHz the detected extragalactic sources are mostly radio loud Active Galactic Nuclei (AGNs), while at frequencies $\ge 353\,$GHz they are mostly dusty galaxies (Planck Collaboration XIII 2011; Planck Collaboration VII 2012). Since we are interested in the latter objects we focus our analysis on the three highest frequency {\it Planck} channels although we could also derive an estimate, over a limited luminosity range, of the luminosity function of dusty galaxies at 217 GHz.

The ERCSC offers, for each source, four different flux density measurements (Planck Collaboration VII 2011). One is that determined by the source detection algorithm (FLUXDET); the others are based on aperture photometry (FLUX), on fitting the source with the {\it Planck} point spread function at the location of the source (PSFFLUX), and on fitting the source with an elliptical Gaussian model (GAUFLUX). As stated in the Explanatory Supplement to the {\it Planck} ERCSC (Planck Collaboration 2011) the ERCSC flux density values need to be corrected by factors depending on the spectral shape of the sources. We have adopted the multiplicative correction factors appropriate for a spectral index $\alpha=3$ ($F_\nu \propto \nu^\alpha$), i.e. 0.896, 0.887, 0.903 and 0.965 at 217, 353, 545 and 857 GHz, respectively, as given in the Explanatory Supplement.

It is important to remember that, in building the ERCSC, the emphasis was more on source reliability than on photometric accuracy or completeness (Planck Collaboration VII 2011), and the absolute calibration of flux-density values was  required to be accurate to within about 30\%, although the signal-to-noise of the sources are much higher. To investigate the best {\it Planck} flux
density measurements and their photometric quality we have exploited {\it Herschel} flux density measurements from the {\it Herschel} Reference Survey (HRS; Boselli et al. 2010), as given in Ciesla et al. (2012; 149 galaxies at 857\,GHz and 86 at 545\,GHz), from the Key Insights on Nearby Galaxies Survey, KINGFISH (Dale et al. 2012, 33 galaxies at 857\,GHz and 33 at 545\,GHz), from the Herschel Virgo Cluster Survey (HeViCS; Davies et al. 2012, 39 galaxies at 857\,GHz and 19 at 545\,GHz), and from the {\it Herschel} Astrophysical Terahertz Large Area Survey (H-ATLAS; Eales et al. 2010) as given in Herranz et al. (2012, 11 galaxies at 857\,GHz and 6 at 545\,GHz). We have applied to the {\it Herschel} flux density measurements colour corrections by factors of 1.004 and 1.02 at 350 and $500\,\mu$m, respectively, appropriate for extended sources with dust temperature of 20\,K and dust emissivity index $\beta=1.5$ (Table 2 of Ciesla et al. 2012).

The {\it Planck} 857 GHz and {\it Herschel} $350\,\mu$m bands have (by design) very similar central wavelengths and the colour corrected flux densities in these bands can be compared directly. The {\it Planck} 545 GHz (=550\,$\mu$m) and {\it Herschel} $500\,\mu$m bands, on the other hand, are slightly offset. In order to compare the measurements in these two nearby bands a small colour correction is needed. In order to do this we extrapolate the {\it Herschel} flux densities to $550\,\mu$m using the spectral index for each source as determined by the measured {\it Planck} flux densities at 857 and 545 GHz. For sources missing {\it Planck} measurements at one of these 2 frequencies we have adopted the average value found for the others, i.e. $\alpha=2.7$. The results of these flux comparisons are shown in Figs.\,\ref{fig:545GHz_comp} and \ref{fig:857GHz_comp}.

In doing this exercise we need to take into account that {\it Planck} measurements may suffer from blending of close-by sources, individually detected by {\it Herschel}. Visually inspecting each HRS galaxy associated to a  {\it Planck} source, Ciesla et al. (2012) found potential contamination of 11 sources out of the 155 in common with {\it Planck} at 350\,$\mu$m. Two {\it Planck} sources in the HeViCS survey, out of 60 observed, are found to be blends of galaxy pairs well resolved by {\it Herschel}: NGC\,4298 $+$ NGC\,4302 and NGC\,4567/8 (Davies et al. 2012). One additional blend of two comparably bright galaxies (NGC\,3719 $+$ NGC\,3720) was found by Herranz et al. (2012), out of 12 detected galaxies in the H-ATLAS equatorial fields.  In these cases, the  {\it Planck} source has been split into its components and we have adopted the {\it Herschel} flux densities of the individual galaxies at 350 and 500\,$\mu$m (the latter rescaled to 550\,$\mu$m). At longer wavelengths we have assigned to the two components flux density values whose sum equals the measured {\it Planck} flux density at that wavelength and whose ratio is equal to the one measured by {\it Herschel} at 500\,$\mu$m. A visual inspection of optical images of the ERCSC sources without {\it Herschel} observations showed that only few per cent of them are associated to  galaxy pairs. Therefore we conclude that source blending is not a major problem for estimates of the luminosity functions using {\it Planck} data.

We have performed a linear fit to the data in Figs.\,\ref{fig:545GHz_comp} and \ref{fig:857GHz_comp}:
\begin{equation}\label{eq:fit}
F_{{\rm Herschel}}=A + B\times F_{{\rm Planck}}.
\end{equation}
The derived coefficients are listed in columns 2 and 3 of Table\,\ref{tab:coefficients_rms}. As the ERCSC is expected to suffer from the Eddington (1913) bias at low flux densities, we have repeated the calculation of the linear fits restricting ourselves to the sub-samples of galaxies with $F_{\rm Planck}>3\,$Jy at 857\,GHz (350\,$\mu$m), and with $F_{\rm Planck}>2\,$Jy at 545\,GHz (550\,$\mu$m); the corresponding coefficients, $A_{\ast}$ and $B_{\ast}$, are also given in Table~\ref{tab:coefficients_rms}. The rms fractional differences between {\it Planck} and  {\it Herschel} flux densities, $\hbox{rms}=(1/\sqrt{N})[\sum_{i=1}^N({F_{\rm Planck,i}-F_{\rm Herschel,i}})^2/F_{\rm Planck,i}^2]^{1/2}$, for the two bright sub-samples are given in the last column of the Table. 

From this analysis we draw the the following conclusions:
\begin{itemize}
\item As expected, all the 4 {\it Planck} measurements are systematically higher than the {\it Herschel} measurements for flux densities below $\lsim 1.5\,$Jy (the best fit values of the coefficient $A$ are all negatives). This is likely due to the Eddington (1913) bias.
\item At 857\,GHz, FLUX is the {\it Planck} flux density measurement most consistent with {\it Herschel} data, as it provides the lowest dispersion around the {\it Herschel} measurements. However, compared to GAUFLUX,  it seems to slightly underestimate the flux density of the brightest fluxes, corresponding to resolved nearby galaxies. This is probably due to the failure of the adopted point source model.
\item At 545\,GHz the {\it Planck} flux density estimates most consistent with {\it Herschel} results are GAUFLUX and FLUXDET, the former showing a slightly lower dispersion of {\it Planck} to {\it Herschel} flux density ratios for the bright sub-sample. Both FLUX and PSFFLUX underpredict the  flux densities of the brightest (i.e. resolved) galaxies, with derived values of the slope $B$ significantly higher than 1.
\end{itemize}
In the light of these considerations, we took FLUX as our reference {\it Planck} flux density measurement at 857\,GHz and used eq.\,(\ref{eq:fit}) with the values of $A$ and $B$ taken from Table\,\ref{tab:coefficients_rms} (first and second columns) to bring it in statistical agreement with {\it Herschel} measurements before deriving the rest-frame luminosities. At 545 GHz we adopted the GAUFLUX values corrected again according to eq.\,(\ref{eq:fit}). At 353\,GHz and 217\,GHz there are no {\it Planck}-independent flux density measurements for a statistically significant number of local galaxies in the {\it Planck} sample (in fact, only a few galaxies in our 353\,GHz {\it Planck} sample have SCUBA imaging data from SLUGS). Since the luminosity function is derived for a sub-sample of relatively bright galaxies (see \S\,2.3), and given that  GAUFLUX seems to perform quite well at both 857 and 545\,GHz for the sub-mm brightest galaxies, we have decided to use that flux density estimate at 353 and 217\,GHz.

\begin{table}
\centering
\footnotesize
\begin{tabular}{lccccc}
\hline
\hline
                                    &                  $A$   &                  $B$      &   $A_{\ast}$   &   $B_{\ast}$    &  rms$_{\ast}$ \\
                                    &                  (Jy)    &                     &         (Jy)                   &                       &                               \\
\hline
\hline
\multicolumn{2}{l}{857\,GHz {\bf (350$\,\mu$m)}} & & & & \\
\hline
FLUX             &    $-$0.4197  &         1.0533     &    $-$0.7933     &    1.0805   &   0.1529 \\
PSFFLUX        &   $-$0.5847   &        1.1645      &   $-$1.1290     &    1.2115   &   0.2121 \\
GAUFLUX     &   $-$0.4073    &         0.9297    &    $-$0.5789     &    0.9326   &  0.1990  \\
FLUXDET      &   $-$0.0876    &        0.9241    &    $-$0.4118     &     0.9405   &   0.1665 \\
\hline
\hline
\multicolumn{2}{l}{545\,GHz {\bf (550$\,\mu$m)}} & & & & \\
\hline
FLUX            &    $-$0.4129  &         1.2904     &    $-$1.0541     &    1.3934   &   0.5927 \\
PSFFLUX       &   $-$0.4660   &        1.3499      &   $-$0.8308     &    1.4079   &   0.7690 \\
GAUFLUX     &   $-$0.2999    &       1.0503    &    $-$0.6107     &   1.0979   &  0.3880  \\
FLUXDET      &   $-$0.0991    &       1.0440    &    $-$0.3830     &     1.1020   &   0.4208 \\
\hline
\hline
\end{tabular}
\caption{Coefficients of the linear least square relation between {\it Planck} and {\it Herschel} flux densities at 857 and 545\,GHz for the whole sample shown in Figs\,\ref{fig:545GHz_comp} and \ref{fig:857GHz_comp} ($A$, $B$) and for the sub-samples with flux density greater than 3\,Jy at 857\,GHz and greater than 2\,Jy at 545\,GHz ($A_{\ast}$ and $B_{\ast}$). Also shown are the rms fractional differences between {\it Planck} and {\it Herschel} flux densities for the bright sub-samples (rms$_{\ast}$).}
\label{tab:coefficients_rms}
\end{table}

\begin{figure*}
\hspace{+1.7cm}
\makebox[\textwidth][c]{
\includegraphics[width=0.87\textwidth]{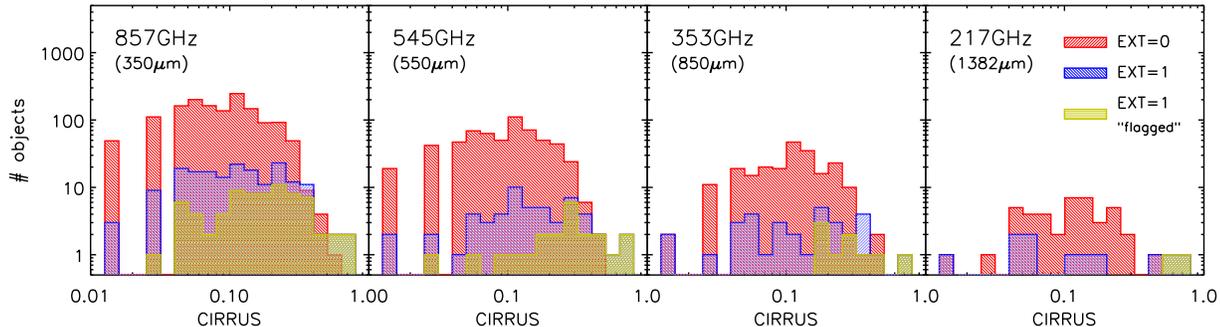}
}
\vspace{-5.2cm}
\caption{Histogram of the flag CIRRUS values for the local galaxies in our sample with EXTENDED=0 (red histogram) and with EXTENDED=1 (blue histogram). The sub-sample of galaxies with EXTENDED=1 that were flagged as ``contaminated by cirrus'' is shown  by the yellow histogram.}
 \label{fig:flag_cirrus}
\end{figure*}

\subsection{Selection of star-forming galaxies}\label{sec:selection}

The first step towards the selection of dusty galaxy samples is the identification and removal of Galactic and AGN-dominated sources. The contamination by Galactic sources was minimized by masking regions heavily affected by Galactic emissions. Planck Collaboration VII (2012) used still unpublished {\it Planck} maps to create the mask. Here we rely instead on the $100\,\mu$m map of \cite{Schlegel1998} that we smoothed with a beam of FWHM$=180^{\prime}$. Pixels in the upper intensity quartile (in the unsmoothed map) are discarded as being contaminated by Galactic emission. The masked region includes 29.1\% of the sky, thus leaving an area of 29,250\,deg$^{2}$ for the measurement of the luminosity function.  The numbers of {\it Planck} detections within that area are listed in Table\,\ref{tab:samples_info}.

In order to help identifying Galactic sources, the ERCSC includes two flags for each source: 'EXTENDED' and 'CIRRUS'. The flag EXTENDED is set to 1 when the square root of the product of the measured major and minor axes of the source is 1.5 times greater than the square root of the product of the major and minor axes of the estimated {\it Planck} point spread function at the location of the source. Otherwise the flag is set to 0. Given the arcmin size of the {\it Planck} beam in the frequency range 857 to 217\,GHz (see \S\,\ref{sect:ERCSC}), detections with EXTENDED=1 are likely to be associated with structure in the Galactic interstellar medium \citep{Herranz2012}. The numbers of {\it Planck} detections with EXTENDED=0  within the selected area are  reported in Table\,\ref{tab:samples_info}, together with that of sources with EXTENDED=1 within the same area. The flag 'CIRRUS' quantifies how crowded the region around each {\it Planck} detection is, by providing the number of sources within a 2$^{\circ}$ radius from the source, in raw 857\,GHz catalogues. The number is normalized to a peak value of unity. The normalization factor is obtained from the number density of sources in the Large Magellanic Cloud region where the highest concentration of 857\,GHz sources is observed. High values of this flag, e.g. CIRRUS $\gsim0.125$ (Herranz et al. 2012), may indicate that the detection has a Galactic origin or, if not, that the flux measurement is severely contaminated by Galactic dust emission. There are exceptions. In fact, some nearby galaxies have either EXTENDED=1 or CIRRUS$>$0.125 or both. One example is M81, which has EXTENDED=1 (at 217\,GHz) and CIRRUS=0.434. Planck Collaboration XVI (2011) found that about 20\% of the objects in their sample of robustly identified nearby galaxies are classified as extended in the ERCSC.
The most effective way to separate local galaxies from Galactic sources would be to visually inspect optical to infrared imaging data for all the {\it Planck} detections. This is the kind of approach we have decided to adopt here. However, as this method is time consuming, we have carried it out on all the detections with EXTENDED=0 and only on a sub-sample with EXTENDED=1 (as only a small fraction of local galaxies are expected to be flagged as extended in the ERCSC). We have decided to not exploit the flag CIRRUS for reasons explained in \S\,\ref{subsub:extended1} where we also describe the definition and the analysis of the sub-sample with EXTENDED=1. We first focus on the catalogue with EXTENDED=0.

\begin{table}
\centering
\begin{tabular}{lcccc}
\hline
\hline
{\bf $\nu_{\rm obs}$ (GHz)}                                &     217    &      353   &   545    &    857     \\
{\bf $\lambda_{\rm obs}$ ($\mu$m)}                                 &     {\bf 1382}  &   {\bf 850}  &  {\bf 550}  &  {\bf 350}     \\
\hline
\hline
$n_{\rm tot}$               &     964    &      1762  &     3781  &        6004   \\
\hline
$n_{\rm EXT=0}$          &      473   &       584  &        946  &        1887   \\
$n_{\rm EXT=1}$          &      491   &     1178  &      2835  &        4117   \\
\hline
$n_{\rm EXT=0, local}$   &            42  &              220   &           598  &     1463  \\
$n_{\rm EXT=1, local}$   &         11(9) &          33(20)   &        55(21)  &      182(56) \\
\hline
\hline
\end{tabular}
\caption{Total number of ERCSC sources, $n_{\rm tot}$, outside the adopted Galactic mask.  Also shown are: the numbers of sources with EXTENDED=0 ($n_{\rm EXT=0}$) and EXTENDED=1 ($n_{\rm EXT=1}$), the number of  identified galaxies at $z<0.1$ with EXTENDED=0 ($n_{\rm EXT=0, local}$) and with EXTENDED=1 ($n_{\rm EXT=1, local}$). In parenthesis are the numbers of the galaxies with EXTENDED=1 and $z\le0.1$ that were kept for estimating the luminosity function (see text for details).}
\label{tab:samples_info}
\end{table}

\subsubsection{Sample with EXTENDED=0}

In order to pick up radio loud AGNs we have exploited the strong difference of their spectral shape compared with that of dusty galaxies at mm and sub-mm wavelengths. In this wavelength range the spectral index ($\alpha$, with $F_{\nu} \propto \nu^{\alpha}$) of radio-loud AGNs is generally $\lsim 0$ while that of dusty galaxies is $\gsim 2.5$. Since the former are increasingly important with decreasing frequency, we have first focused on the 217\,GHz sample and cross-correlated it with the ERCSC at 100\,GHz in order to estimate the 100-to-217\,GHz spectral index. We found 309 (out of 473) sources with $F_{100}/F_{217}>1$ and dropped them from the sample as being dominated by radio emission. Note that this approach is different from that adopted by Planck Collaboration VII (2012), where the 857-to-545\,GHz and 545-to-353\,GHz flux density ratios were used to separate radio sources from galaxies whose spectral energy distribution is dominated by thermal dust emission.

As a counter-check, we cross-correlated our 217\,GHz sample with the CRATES catalogue (Healey et al. 2007) adopting a search radius of $3^{\prime}$. This radius is somewhat larger, for these {\it Planck} channels, than
the 1/2 FWHM of the {\it Planck} beam found by Planck Collaboration XIV (2011) to be sufficient to locate any related source. CRATES provides a nearly uniform extragalactic ($|b| > 10^\circ$) coverage for flat-spectrum ($\alpha > -0.5$) radio sources brighter than $65\,$mJy at 4.8\,GHz. As expected, since radio sources detected by {\it Planck} are generally flat-spectrum, most of the sources dropped as synchrotron dominated have a bright CRATES counterpart. The cross-correlation with CRATES however misses a bunch of very bright steep-spectrum sources (e.g. 3C\,207, 3C\,216, 3C\,286, 3C\,380) and includes some sources whose (sub)-mm emission is clearly dominated by dust (e.g. M82, M83, M104). On the whole, the cross-correlation with CRATES confirmed our previous conclusions and did not reveal any misclassified source.

The remaining $473-309=164$ sources were inspected individually.  We used the  interactive software sky atlas Aladin (Bonnarel et al. 2000) to inspect the available optical to far-infrared imaging data (e.g. IRAS maps) and to query the NASA/IPAC Extragalactic Database (NED)\footnote{http://ned.ipac.caltech.edu/} around the position of each source. We removed all those objects that did not have a nearby ($z<0.1$) galaxy  within the adopted search radius but, instead,  satisfied one (or more) of the following conditions: (i) association with an extended clumpy/filamentary structure in the IRAS maps (if available); (ii) presence of a bright (i.e. $F_{\rm 1.4GHz}>0.1$\,Jy) radio source; (iii)  presence of a group/cluster of galaxies (examples are Abell 2218 and the Bullet cluster); (iv) the {\it Planck} sources is un-detected in IRAS maps. The last condition is meant to eventually identify and remove false {\it Planck} detections that may have been induced by background fluctuations, which are expected to be significant in low-resolution surveys (see e.g. Negrello et al. 2004, 2005).  We identified 122 objects obeying at least one of the conditions listed above (we will refer to them as ``contaminants'').
Removing those sources from the catalogue left us with a final, cleaned, sample of 42 local galaxies at 217\,GHz with EXTENDED=0, all being either Messier objects or galaxies listed in the New General Catalogue (NGC).
We then moved to the 353\,GHz sample and cross-correlated it with the 217\,GHz catalogue, using a 3$^{\prime}$ search radius. We removed all the ``contaminants'' in common with the 217\,GHz sample and kept all the sources that were identified  as local galaxies at that frequency. We then inspected each of the remaining unclassified {\it Planck} detections  individually, in Aladin, and removed from the sample all those objects satisfying at least one of the conditions previously illustrated. We followed the same approach at 545 and at 857\,GHz after cross-correlating the catalogues with the lower frequency samples in order to exploit all the information already in hand.

The whole cleaning process, performed on the catalogue of {\it Planck} detections with EXTENDED=0, produced a sample of 220 local galaxies at 353\,GHz, 598 at 545\,GHz, and 1463 at 857\,GHz (see Table\,\ref{tab:samples_info}), the majority of which are either Messier or NGC objects.
The visual inspection of multi-wavelength ancillary data for each individual object minimized the risk of misidentifications and automatically provided, through the NED, the information on the redshift-independent measurements of distance, needed to derive the local luminosity function.

\begin{figure}
\vspace{0.0cm}
\hspace{-4.5cm}
\makebox[\textwidth][c]{
\includegraphics[width=0.70\textwidth]{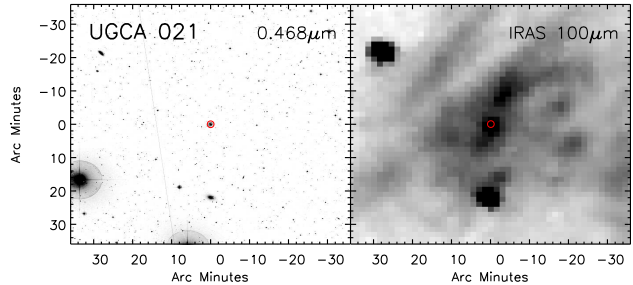}
}
\vspace{-4.0cm}
\caption{Postage stamp image of the galaxy UGCA-021 ($z=0.006622$), at 0.468$\,\mu$m (left-hand panel) and at 100\,$\mu$m (right-hand panel). This galaxy, indicated by the red circle, is flagged as extended in the ERCSC at 545\,GHz and has CIRRUS=0.094. However the IRAS map shows that it lies behind a prominent structure in the Galactic interstellar medium.}
 \label{fig:UGCA-021}
\end{figure}

\begin{figure}
\vspace{0.5cm}
\hspace{-6.5cm}
\makebox[\textwidth][c]{
\includegraphics[width=0.7\textwidth]{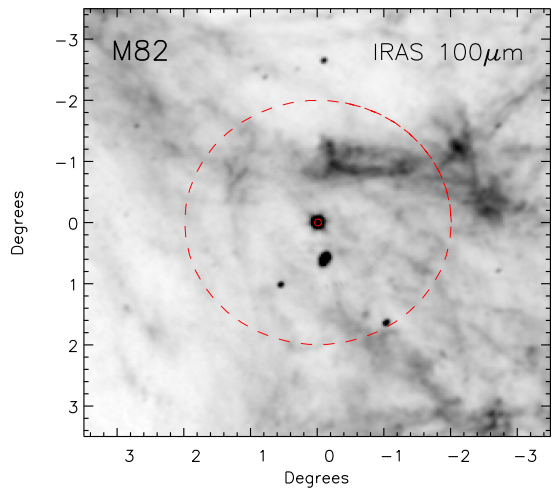}
}
\vspace{-1.0cm}
\caption{Postage stamp image of the galaxy M82 ($z=0.00068$), at 100\,$\mu$m. The galaxy, indicated by the red circle, has EXTENDED=0 at 545\,GHz and CIRRUS=0.5. The dashed red circle marks the region of 2$^{\circ}$ radius centered on the galaxy. The high value of the flag CIRRUS is likely due to the presence, in the vicinity, of a prominent structure in the Galactic interstellar medium, seen in the top-right corner of the image.}
 \label{fig:M82}
\end{figure}

\subsubsection{Sample with EXTENDED=1}\label{subsub:extended1}

Table\,\ref{tab:samples_info} shows that, particularly at 857\,GHz and 545\,GHz, the {\it Planck} detections with EXTENDED=1 are a factor $\gsim$2 more abundant than those with EXTENDED=0. Visually inspecting all those sources would be a very inefficient method to identify local galaxies, as the majority of the {\it Planck} detections with EXTENDED=1 are expected to be Galactic. Therefore we have first reduced the sample by performing an automatic cross-matching with the NED and by retaining for inspection only the {\it Planck} detections that happen to have at least one source with a measured redshift $z<0.1$ within a distance of 3$^{\prime}$.  A galaxy like the Milky Way, with a linear size of $\sim$30\,kpc, would be assigned EXTENDED=1 for distances $\lsim20\,$Mpc or, equivalently, for redshifts $z\lsim0.005$. Therefore our choice of an upper limit of 0.1 in redshift is very conservative. The numbers of objects left over after the cross-matching with the NED are reported in Table\,\ref{tab:samples_info}. Upon visual inspection of each individual source in Aladin, to identify and remove radio sources and Galactic objects, we flagged as ``contaminated by cirrus'' those galaxies that happened to lie behind a prominent filamentary structure in the IRAS 100\,$\mu$m map. These galaxies are likely to have their {\it Planck} flux density measurements severely affected by Galactic emission and therefore we have decided to drop them from the final sample. The number of galaxies with EXTENDED=1 actually used for deriving the luminosity function is shown in parenthesis in Table\,\ref{tab:samples_info}. One might have used the flag CIRRUS to identify local galaxies falling behind  prominent Galactic structures, by imposing an upper limit on the CIRRUS values, for example 0.125, as suggested by Herranz et al. (2012). However Fig.\,\ref{fig:flag_cirrus} demonstrates that this approach is prone to misidentifications. In fact, the figure shows the histogram of the flag CIRRUS values for the sample of  local galaxies with EXTENDED=0 (in red) and that with EXTENDED=1 (in blue), while the yellow histogram refers to the sub-sample of galaxies with EXTENDED=1 that were  flagged as ``contaminated by cirrus''.  While the latter objects are indeed those  with the highest CIRRUS values at 217\,GHz and 353\,GHz, the situation becomes less well defined at shorter wavelengths where several of them turn out to have CIRRUS$\lsim$0.1. One example is shown in Fig.\,\ref{fig:UGCA-021}.
On the other hand there are some nearby galaxies that, upon inspection of the IRAS 100$\,\mu$m maps, do not seem to be contaminated by Galactic cirrus but, nevertheless, have CIRRUS$>$0.2. Examples are M64, M82, M90, NGC7392. As illustrated in Fig.\,\ref{fig:M82} for M82, the high CIRRUS value in these sources is probably due to the presence, in the vicinity of the galaxy (i.e. within 2$^{\circ}$) but not directly on top of it, of a prominent structure in the Galactic interstellar medium.

\begin{table}
\centering
\begin{tabular}{cccccc}
\hline\hline
$\nu_{\rm obs}$  &  {\bf $\lambda_{\rm obs}$} & $n_{\rm obj}$  & $F_{\rm lim}$ & $\log\bar{F}$ & $\sigma_{log F}$\\
(GHz) & ($\mu$m) &  &  (mJy)  &  &  \\
\hline
857   &  {\bf 350}  &  328  &  4086 &  2.71  &  0.95 \\ 
545   &  {\bf 550}  &  234  &  1814 &  2.63  &  0.66 \\ 
353   &  {\bf 850}  &  108  &    809  &  2.39  &  0.55 \\ 
217  &   {\bf 1382}  &  30  &     497  &  2.27  &  0.44 \\ 
\hline
\hline
\end{tabular}
\caption{Flux density limits ($F_{\rm lim}$) corresponding to 80\% completeness and numbers of local dusty galaxies brighter than these limits, used for estimating the luminosity functions. The quantities $\log\bar{F}$ and $\sigma_{\log F}$ are the best-fit values of the parameters that define the completeness function [eq.~(\protect\ref{eq:completeness})], where the flux density is in mJy.}
\label{tab:info}
\end{table}

\begin{figure*}
\hspace{+1.7cm}
\makebox[\textwidth][c]{
\includegraphics[width=0.87\textwidth]{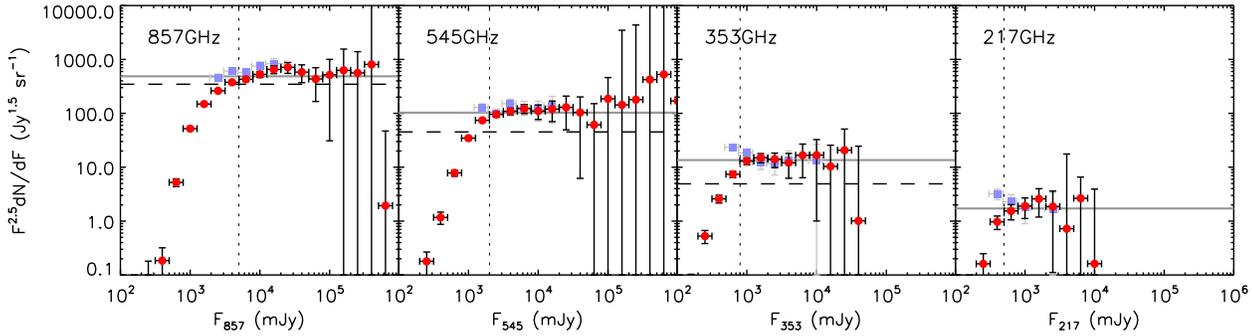}}
\vspace{-5.2cm}
\caption{Euclidean normalized number counts of local star-forming galaxies at 857\,GHz {\bf (350$\,\mu$m)}, 545\,GHz {\bf (550$\,\mu$m)}, 353\,GHz {\bf (850$\,\mu$m)} and 217\,GHz {\bf (1382$\,\mu$m)} (red dots with error bars). Also shown, for comparison, are the counts estimated by Planck Collaboration VII (2012; blue squares,  corrected for incompleteness), as well as the predictions by Serjeant \& Harrison (2005; dashed lines). The thick grey line is the result of the fit to the data points with flux density above the value indicated by the vertical dotted line, assuming that the differential counts have an Euclidean slope.}
 \label{fig:dNdlgF_euc}
\end{figure*}

\begin{figure*}
\hspace{+1.7cm}
\makebox[\textwidth][c]{
\includegraphics[width=0.87\textwidth]{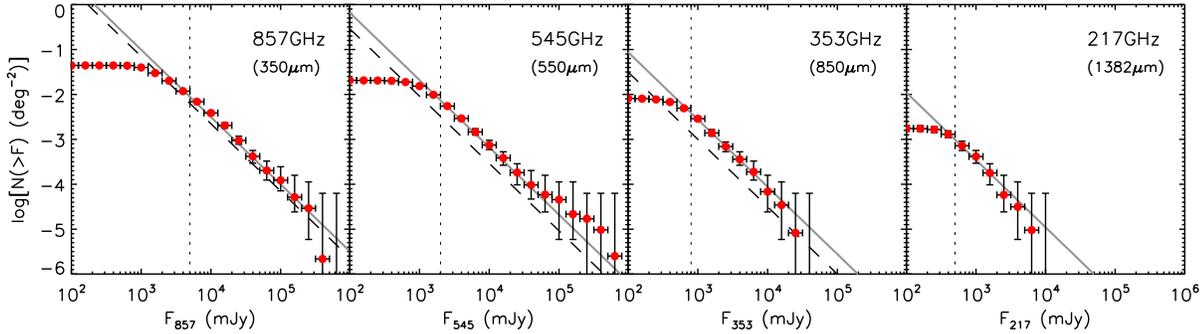}}
\vspace{-5.2cm}
\caption{ Integral number counts of local star-forming galaxies at 857\,GHz (350$\,\mu$m), 545\,GHz (550$\,\mu$m), 353\,GHz (850$\,\mu$m) and 217\,GHz (1382$\,\mu$m) (red dots with error bars). The dashed line is the prediction by Serjeant \& Harrison (2005). The thick grey line represents the integral number counts derived from the fit to the bright tail of the differential counts shown in Fig.\,\ref{fig:dNdlgF_euc}.}
 \label{fig:NgtF}
\end{figure*}

\subsection{Number counts and completeness}\label{subsec:diff_counts}

\begin{table*}
\centering
\begin{tabular}{rrrrr}
\hline\hline
$\log[S \rm (mJy)]$         & $(S^{2.5}dN/dS)_{\rm 857}$  &  $(S^{2.5}dN/dS)_{\rm 545}$  &  $(S^{2.5}dN/dS)_{\rm 353}$  &  $(S^{2.5}dN/dS)_{\rm 217}$   \\
                                       & (Jy$^{1.5}$\,sr$^{-1}$)  &  (Jy$^{1.5}$\,sr$^{-1}$)  &   (Jy$^{1.5}$\,sr$^{-1}$)  &   (Jy$^{1.5}$\,sr$^{-1}$)   \\
\hline
       2.8    &        -                                  &         -                                             &         -                                  &       1.6$\pm$0.5    \\
       3.0    &        -                                  &         -                                             &         13.0$\pm$1.9               &       1.9$\pm$0.7     \\
       3.2    &        -                                  &         -                                              &         15.0$\pm$2.9             &       2.6$\pm$1.4    \\
       3.4    &        -                                  &         96$\pm$10                            &         14.0$\pm$4.2               &       1.9$\pm$1.8         \\
       3.6    &        -                                  &         109$\pm$15                          &         12.3$\pm$6.0               &        0.7$\pm$17.2          \\
       3.8    &        429$\pm$41               &         121$\pm$24                           &         17$\pm$10               &       2.8$\pm$4.1     \\
       4.0    &        528$\pm$67               &         109$\pm$33                           &         17$\pm$16               &         -                                  \\
       4.2    &        652$\pm$108              &        121$\pm$50                             &         10$\pm$15                   &         -                                  \\
       4.4    &        715$\pm$164             &         126$\pm$78                             &         19$\pm$29                &         -                                  \\
       4.6    &        582$\pm$212              &        103$\pm$97                          &         -                                  &         -                                  \\
       4.8    &        434$\pm$268              &        57$\pm$1336                         &         -                                  &         -                                  \\
       5.0    &        517$\pm$486              &        184$\pm$272                        &         -                                  &         -                                  \\
       5.2    &        630$\pm$929            &         -                                           &         -                                  &         -                                  \\
       5.4    &        563$\pm$830               &         -                                              &         -                                  &         -                                  \\
       5.6    &        -                                  &         -                                             &         -                                  &         -                                  \\
\hline
\end{tabular}
\caption{Euclidean normalized number counts of local star-forming galaxies at 857\,GHz {\bf (350$\,\mu$m)}, 545\,GHz {\bf (550$\,\mu$m)}, 353\,GHz {\bf (850$\,\mu$m)} and 217\,GHz (1382$\,\mu$m), for flux densities greater than the values indicated by the vertical dotted lines in Figs.~\ref{fig:dNdlgF_euc} and \ref{fig:NgtF}.}
\label{tab:dNdlgF}
\end{table*}

The completeness of our samples of local dusty galaxies can be tested by means of the differential source counts. Apart from the effect of inhomogeneities in the spatial distribution of galaxies, the counts must have an Euclidean slope (i.e. $dN/dS\propto S^{-2.5}$) and the onset of incompleteness is indicated by a decline below the Euclidean power law. The counts were estimated using a bootstrap resampling method that allows us to account for the uncertainties in the flux density measurements. In practise we have generated 1000 simulated catalogues by resampling, with repetitions, the input catalogue. In each simulation we have assigned to each source a flux density value randomly generated from a Gaussian probability distribution with a mean equal to the measured flux density and dispersion $\sigma$ equal to the associated error. The distribution of source flux densities derived from the simulations were binned into a histogram and the mean value in each bin was taken as the measurement of the number count in that bin, once divided by the bin size and by the survey area.  The errors on the number counts were derived assuming a Poisson statistic, according to the prescriptions of Gehrels (1986).

The results are shown in Fig.\,\ref{fig:dNdlgF_euc} in the form of Euclidean normalized differential number counts and in Fig\,\ref{fig:NgtF} as integral number counts. As expected, the counts exhibit an Euclidean slope at the brightest flux densities, followed by a downturn (or a flattening in the cumulative counts), which is due to the onset of incompleteness in the sample. The counts are compared with those estimated for dust-dominated galaxies by Planck Collaboration VII (2012), who used (i) a Galactic mask derived from {\it Planck} data, (ii) an automatic procedure to identify dusty galaxies based on {\it Planck} colours, (iii) completeness corrections obtained from extensive simulations, (iv) FLUX as the reference flux density measurement at all frequencies. The agreement is generally good, particularly at 857 and 545\,GHz. At 353\,GHz and at 217\,GHz the faintest counts produced by the {\it Planck} collaboration lie above the extrapolation  of our measurements. This may be due to the appearance of a strongly evolving population with a ``super-Euclidean'' slope. However the possibility of either an overestimate of the correction for incompleteness or of a contamination of the sample e.g. by Galactic and/or radio sources can hardly be ruled out at this stage.

Fitting the brightest part of the Euclidean normalized counts (i.e. above 5, 2, 0.8, and 0.5\,Jy at 857, 545, 353 and 217\,GHz, respectively; these flux density limits are indicated by the vertical dotted lines in Figs.\,\ref{fig:dNdlgF_euc} and \ref{fig:NgtF}) with a straight horizontal line we get 486$\pm$32, 103$\pm$8, 14$\pm$1 and 1.7$\pm$0.4\,Jy$^{1.5}$\,sr$^{-1}$ at 857, 545, 353 and 217\,GHz, respectively. These values are fully consistent with those estimated by Planck Collaboration VII (2012) for dusty galaxies: 627$\pm$152, 125$\pm$15, and 15$\pm$6 at 857, 545 and 353\,GHz, respectively (at 217\,GHz the {\it Planck} collaboration provides the Euclidean plateau level for synchrotron dominated galaxies only).

The completeness as a function of the flux density was estimated as the ratio of the observed counts to those expected from the best-fit Euclidean counts. It is well approximated by an error function
\begin{equation}\label{eq:completeness}
C(F) = {\rm erf}[(\log{F}-\log\bar{F})^{2}/\sigma_{\log F}^{2}],
\end{equation}
where $\log\bar{F}$ and $\sigma_{\log F}$ are free parameters whose best fit values for each {\it Planck} channel are given in Table~\ref{tab:info}.
For the estimation of the luminosity functions we adopted the flux density limits corresponding to an 80\% completeness.  These limits are listed in Table~\ref{tab:info} and represent {\it observed} flux densities, i.e. no correction for either CO-line emission (see next sub-section) or Eddington-bias [i.e. eq.\,(\ref{eq:fit})] was applied when measuring the number counts. In fact those corrections are only introduced when calculating the rest-frame luminosities. In the same table we also give the corresponding numbers of local dusty galaxies used to estimate the luminosity functions.  All of them have a spectroscopic redshift. The completeness curves [eq.~(\ref{eq:completeness})] are also used to compute the weight factor for each source, $w_{i}=1/C(F_{i})$, $F_{i}$ being the flux of the $i$-$th$ galaxy, to be used in the estimate of the luminosity function [see eq.\,(\ref{eq:LF_Vmax}) and eq.\,(\ref{eq:completeness})].

By comparing our number counts with those predicted by Serjeant \& Harrison (dashed lines in Figs.~\protect\ref{fig:dNdlgF_euc} and \ref{fig:NgtF}), we find that the latter lie significantly below the former, particularly at the longest wavelengths, implying that the Serjeant \& Harrison local luminosity functions were also underestimated. This is not surprising since the Serjeant \& Harrison estimates were obtained extrapolating the IRAS 60$\,\mu$m data to 850$\,\mu$m using the sub-millimeter/far-infrared colour relation derived from the SLUGS \citep{Dunne00}. The latter is known to be biased against galaxies with
large cold dust components (see Vlahakis et al. 2005 and the discussion in \ref{sec:results}), which are instead readily detected by {\it Planck} (Planck Collaboration XVI 2011).

\subsection{Correction of {\it Planck} fluxes for CO line emission}

All the {\it Planck} High Frequency Instrument (HFI) bands, except that at $143\,\rm GHz$, include emission from the $CO$ molecule in low redshift galaxies. This emission is a contaminant for our purposes as we are concerned with the dust continuum emission from star forming galaxies. Since most of the sources in our samples lack $CO$ data and we are dealing with the properties of the population as a whole, we apply to the {\it Planck} fluxes average correction factors for contamination from the various CO lines before deriving the corresponding rest-frame luminosities.

The correction factors are based on the correlation between the total far-infrared luminosity ($L_{\rm FIR}$) and the $CO$ line emission, $L_{CO}$, and specifically on the correlation presented by Genzel et al. (2010) who found a close
to linear relation between $L_{CO}$ and $L_{\rm FIR}(50\hbox{--}300\mu{\rm m})$ for normal star-forming galaxies with an average ratio $L_{\rm FIR}/L'_{CO(1-0)} = 27\,L_\odot/(\hbox{K}\,\hbox{km}\,\hbox{s}^{-1}\,\hbox{pc}^2)$. The units of $L'_{CO(1-0)}$, ($\hbox{K}\,\hbox{km}\,\hbox{s}^{-1}\,\hbox{pc}^2$) can be converted to solar luminosities with the relation given by Solomon et al (1997), $L_{CO}/L_\odot = 3.2\times 10^{-11} \,(\nu_{\rm rest}/{\rm GHz})^3 \,(L'_{CO}/(\hbox{K}\,\hbox{km}\,\hbox{s}^{-1}\,\hbox{pc}^2)$, to find $L_{\rm FIR}/L_{CO(1-0)}=5.55 \times 10^5$ with both luminosities in solar units.  Although there is evidence that the correlation is different for merger systems with high values of $L_{\rm FIR}$ (see e.g. Daddi et al. 2010) we did not feel that there were enough such systems in our sample to justify refinement of the corrections.

In order to correct the {\it Planck} fluxes contaminated by $CO(J=2-1)$ ($217\,\rm GHz$ band), $CO (J=3-2)$ ($353\,\rm GHz$ band), and $CO (J=5-4)$ ($545\,\rm GHz$ band)\footnote{No correction was made for the $CO(J=7-6)$ and $(J=8-7)$ lines that contaminate the $857\,\rm GHz$   band as their estimated contribution is negligible.}, we assumed fixed line strength ratios. The $CO(J=2-1)/(J=1-0)$ ratio in antenna temperature units, $r_{21}$, was taken as 0.8, which is a value typical of nearby galaxies (Leroy et al. 2009). The $CO(J=3-2)/(J=1-0)$ ratio, $r_{31}$, was taken as 0.66 based on the results of Yao et al. (2003). A similar {\it median} value, $r_{31, \rm median}\simeq 0.7$, was found by Mao et al. (2012; their mean value is however somewhat higher, $r_{31, \rm mean}\simeq 0.81$), while Papadopoulos et al. (2011) find $r_{31, \rm mean}\simeq 0.49$. Finally,  $r_{51}=CO(J=5-4)/(J=1-0)$ was taken as 0.3 (Papadopoulos et al. 2011).

The procedure adopted was as follows. The IRAS 60 and $100\,\rm \mu m$ fluxes (available for all our sources) were used to calculate the FIR ($50\hbox{--}300\mu{\rm m}$) luminosity, $L_{\rm FIR}$, in the same way as done by Genzel et al. (2010). This gives a prediction for the CO (J=1-0) luminosity. This was then scaled by the line ratios specified above, converted to flux units [$F_{CO(J,J-1)}/F_{CO(1-0)}= r_{J1}(\nu_{J,J-1}/\nu_{1-0})^2$], to give the luminosity in the various other $CO$  transitions. The measured {\it Planck} flux density was then used with the predicted $CO$ flux to calculate the equivalent width of the CO line. The ratio of this width and the bandwidth of the {\it Planck} band (taken to be 30\% of the centre frequency, as indicated in the ERCSC Explanatory Supplement) was used as a measure of the fractional contribution of the CO line to the Planck flux. The average correction factors for CO contamination were 0.91, 0.97 and 0.99 at, 217, 357 and 545 GHz, respectively. These values are consistent with the level of contamination estimated by Planck Collaboration XVI (2011).

\begin{figure*}
\vspace{-3.5cm}
\hspace{+0.0cm}
\makebox[\textwidth][c]{
\includegraphics[width=0.95\textwidth]{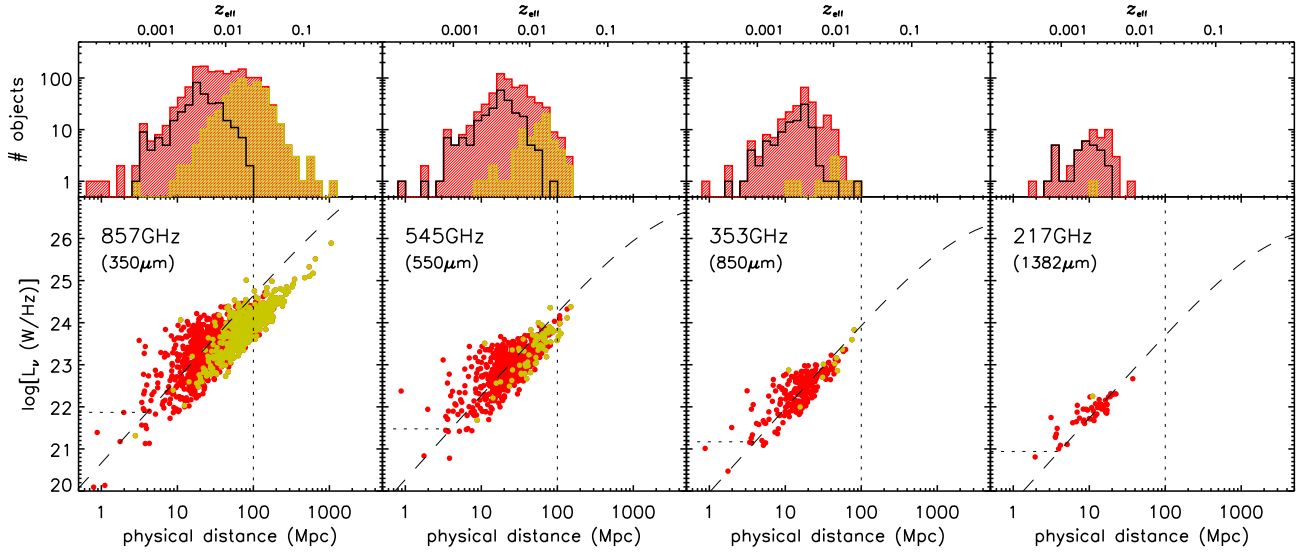} }
\vspace{+0.5cm}
\caption{{\it Top panels}: distributions of source distances at 857\,GHz (350$\,\mu$m), 545\,GHz (550$\,\mu$m), 353\,GHz (850$\,\mu$m) and 217\,GHz (1382$\,\mu$m) for the total samples of local star-forming galaxies (red histograms), for the sub-sample without redshift-independent distance measurements (yellow histograms) and for the sub-sample with completeness above 80\,\% and luminosity higher than the adopted minimum value (black line). The redshift corresponding to a given distance as determined by the cosmic Hubble flow alone (i.e. the one derived by neglecting peculiar motion, that we refer to as {\it effective} redshift, $z_{\rm eff}$) is shown in the upper x-axis. {\it Bottom panels}: monochromatic luminosity as a function of distance at 857, 545, 353 and 217\,GHz. The colour code is the same as in the top panels. The dashed line shows the luminosity that corresponds to the adopted flux density limit as a function of distance, while the horizontal dotted line indicates the adopted minimum luminosity. Galaxies in the samples used to compute the luminosity functions have distances $\simlt 100\,$Mpc (a limit indicated by the vertical dashed line).}
\label{fig:Nz_Lrf_dist}
\end{figure*}
%

\begin{figure*}
\vspace{+0.6cm}
\hspace{+0.0cm}
\makebox[\textwidth][c]{
\includegraphics[width=0.95\textwidth]{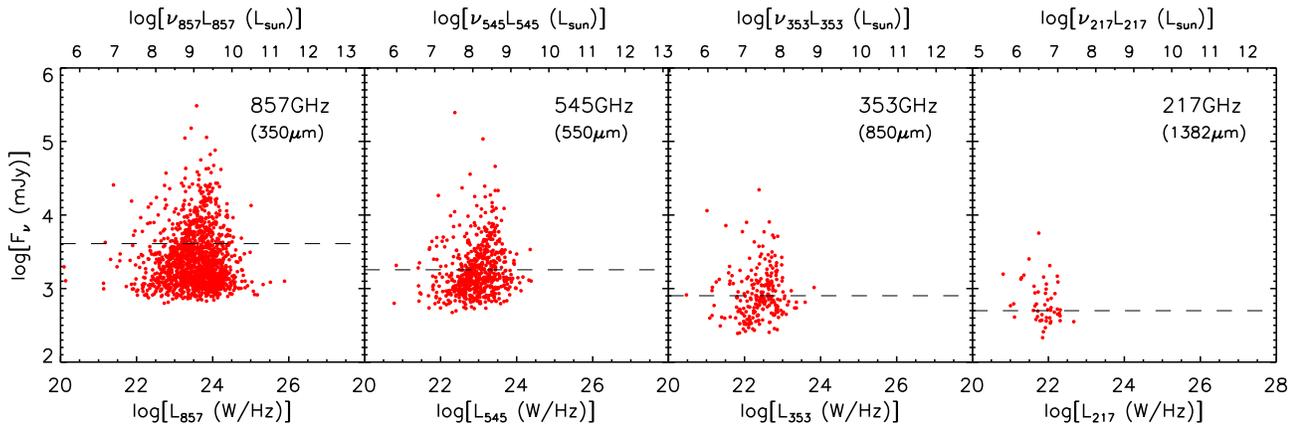} }
\vspace{-5.7cm}
\caption{Measured flux density as a function of the estimated monochromatic luminosity at 857\,GHz (350$\,\mu$m), 545\,GHz (550$\,\mu$m), 353\,GHz (850$\,\mu$m) and 217\,GHz (1382$\,\mu$m). The horizontal dashed line indicates the flux limit above which the sample is 80\% complete (see \S\,\ref{subsec:diff_counts}).}
\label{fig:F_vs_Lrf}
\end{figure*}

\subsection{Distances and rest-frame luminosities}

Although a spectroscopic redshift is available for all the galaxies in our samples, this does not provide, in general, an accurate estimate of the source distance, as in the local Universe  peculiar motions may introduce significant extra red- or blue-shifts in galaxy spectra. Redshift-independent distance estimates should therefore be used as far as possible. Such estimates are available in the NED for the majority of the galaxies in our samples: 65\% of them at 857\,GHz, 88\% at 545\,GHz, 95\% at 353\,GHz and 98\% at 217\,GHz. For the remaining galaxies, distances were computed from the redshifts, previously corrected to the reference frame defined by the Cosmic Microwave Background radiation using the NED on-line calculator\footnote{http://ned.ipac.caltech.edu/help/velc$\underline{~}$help.html}. We assigned to the redshift-dependent distances an arbitrary error of 30\%. The distribution of the source distances at each frequency, shown in the top panels of Fig.\,\ref{fig:Nz_Lrf_dist}, illustrates that these are truly local samples: the distances for the flux-limited sample used to estimate the luminosity functions are all $\le 100\,$Mpc (i.e. $z\lsim0.02$). \\
The rest-frame luminosity of the $i-$th galaxy  at the frequency of observation $\nu_{\rm obs}$  is given by
\begin{equation}
L_{ i} = \frac{4\pi d_{i}^{2}}{(1+z_{i})k(z_{i})}F_{i},
\end{equation}
where $F_{i}$ is the source flux at $\nu=\nu_{\rm obs}$, $d_{i}$ is the source distance, $z_{i}$ its redshift and $k(z_{i})=L_(\nu_{\rm obs}(1+z_{i}))/L(\nu_{\rm obs})$ is the $k-$correction. The latter is estimated using the SED of an Sd galaxy taken from the SWIRE template library\footnote{{\small www.iasf-milano.inaf.it/$\sim$polletta/templates/swire$\underline{~}$templates .html}} (Polletta et al. 2007).
Figure~\ref{fig:F_vs_Lrf} shows the distribution of flux densities as a function of source luminosity. This is meant to illustrate the luminosity range covered by the samples of sources above the adopted flux density limit and how the luminosity bins are populated. The distribution of luminosities as a function of the source distance is shown in the lower panels of Fig.\,\ref{fig:Nz_Lrf_dist}. In practice,  the luminosity function is estimated from galaxies within a distance of 100\,Mpc, as indicated by the vertical dashed line in Fig.\,\ref{fig:Nz_Lrf_dist}.

\begin{table*}
\centering
\begin{tabular}{cccc|cc|cc}
\hline\hline
${\rm log}(L_{857})$  &  ${\rm log}(\phi_{857})$ &  ${\rm log}(L_{545})$  &  ${\rm log}(\phi_{545})$ &  ${\rm log}(L_{353})$  &  ${\rm log}(\phi_{353})$ &  ${\rm log}(L_{217})$  &  ${\rm log}(\phi_{217})$ \\
\hline
22.02  &  $-$1.32$_{-0.32}^{+0.18}$  &   21.63  & $-$1.29$_{-0.31}^{+0.18}$  &   21.32  &  $-$1.37$_{-0.33}^{+0.19}$  &   21.09  &  $-$1.52$_{-0.44}^{+0.21}$  \\
22.32  &  $-$1.55$_{-0.21}^{+0.14}$  &   21.93  & $-$1.47$_{-0.20}^{+0.14}$  &   21.62  &  $-$1.62$_{-0.24}^{+0.15}$  &   21.39  &  $-$1.85$_{-0.29}^{+0.17}$  \\
22.62  &  $-$1.75$_{-0.17}^{+0.12}$  &   22.23  & $-$1.73$_{-0.16}^{+0.12}$  &   21.92  &  $-$1.89$_{-0.19}^{+0.13}$  &   21.69  &  $-$2.07$_{-0.23}^{+0.15}$  \\
22.92  &  $-$1.94$_{-0.10}^{+0.08}$  &   22.53  & $-$1.95$_{-0.10}^{+0.08}$  &   22.22  &  $-$2.03$_{-0.11}^{+0.09}$  &   21.99  &  $-$2.42$_{-0.18}^{+0.13}$  \\
23.22  &  $-$1.99$_{-0.06}^{+0.06}$  &   22.83  & $-$2.02$_{-0.06}^{+0.06}$  &   22.52  &  $-$2.25$_{-0.08}^{+0.07}$  &   22.29  &  $-$2.99$_{-0.24}^{+0.16}$  \\
23.52  &  $-$2.24$_{-0.05}^{+0.04}$  &   23.13  & $-$2.32$_{-0.06}^{+0.05}$  &   22.82  &  $-$2.75$_{-0.09}^{+0.08}$  &              &                                         \\
23.82  &  $-$2.71$_{-0.05}^{+0.04}$  &   23.43  & $-$2.86$_{-0.06}^{+0.06}$  &   22.12  &  $-$3.66$_{-0.19}^{+0.13}$  &              &                                     \\
24.12  &  $-$3.37$_{-0.07}^{+0.06}$  &   23.73  & $-$3.68$_{-0.11}^{+0.09}$  &   23.42  &  $-$4.76$_{-0.58}^{+0.24}$  &              &                                      \\
24.42  &  $-$4.17$_{-0.11}^{+0.09}$  &   24.03  & $-$4.60$_{-0.20}^{+0.14}$  &   23.72  &  $-$5.79$_{-5.00}^{+0.37}$  &              &                                      \\
24.72  &  $-$5.22$_{-0.31}^{+0.18}$  &   24.33  & $-$5.74$_{-0.92}^{+0.27}$  &              &                                          &              &                                      \\
\hline
\hline
\end{tabular}
\caption{Local luminosity function at 857\,GHz (350$\,\mu$m), 545\,GHz (550$\,\mu$m), 353\,GHz (850$\,\mu$m) and 217\,GHz (1382$\,\mu$m) measured via the $1/V_{\rm max}$ method, in units of Mpc$^{-3}\hbox{dex}^{-1}$. Luminosities are in units of W/Hz.}
\label{tab:LF_tabulated}
\end{table*}

\begin{figure*}
\vspace{0.5cm}
\hspace{+0.0cm}
\makebox[\textwidth][c]{
\includegraphics[width=0.95\textwidth]{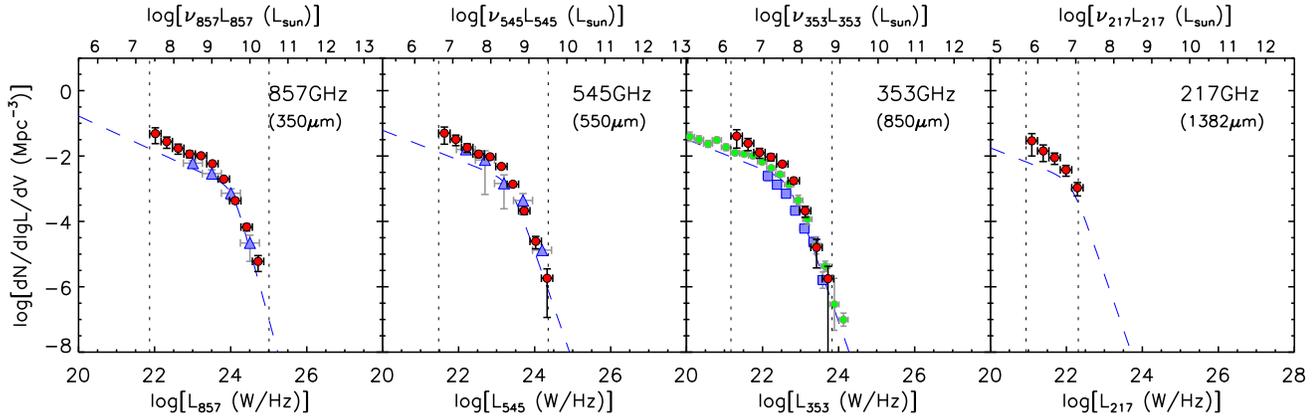}
}
\vspace{-5.7cm}
\caption{Local luminosity functions at 857\,GHz (350$\,\mu$m), 545\,GHz (550$\,\mu$m), 353\,GHz (850$\,\mu$m) and 217\,GHz (1382$\,\mu$m) estimated with the $1/V_{\rm max}$ method.  Also shown, for comparison, are the LF measured by Dunne et al. (2000) and Vlahakis et al. (2005) at 353\,GHz (850\,$\mu$m; blue squares and green squares, respectively), and by Vaccari et al. (2010) at 350$\,\mu$m and at 500$\,\mu$m (blue triangles). The {\it Herschel} luminosities have been converted from 500 to 550\,$\mu$m by assuming a spectral index $\alpha=2.7$, the mean value found for local galaxies in the {\it Planck} sample. The dashed curves at 857, 545 and 353\,GHz represent the predictions by Serjeant $\&$ Harrison (2005). The dashed curve at 217\,GHz is the extrapolation of the Serjeant \& Harrison prediction from 353\,GHz assuming again a spectral index $\alpha=2.7$. The vertical dotted lines correspond to the adopted minimum and maximum luminosities (see text).}
 \label{fig:lf_Vmax}
\end{figure*}
%

\begin{figure*}
\vspace{+0.5cm}
\hspace{+0.0cm}
\makebox[\textwidth][c]{
\includegraphics[width=0.95\textwidth]{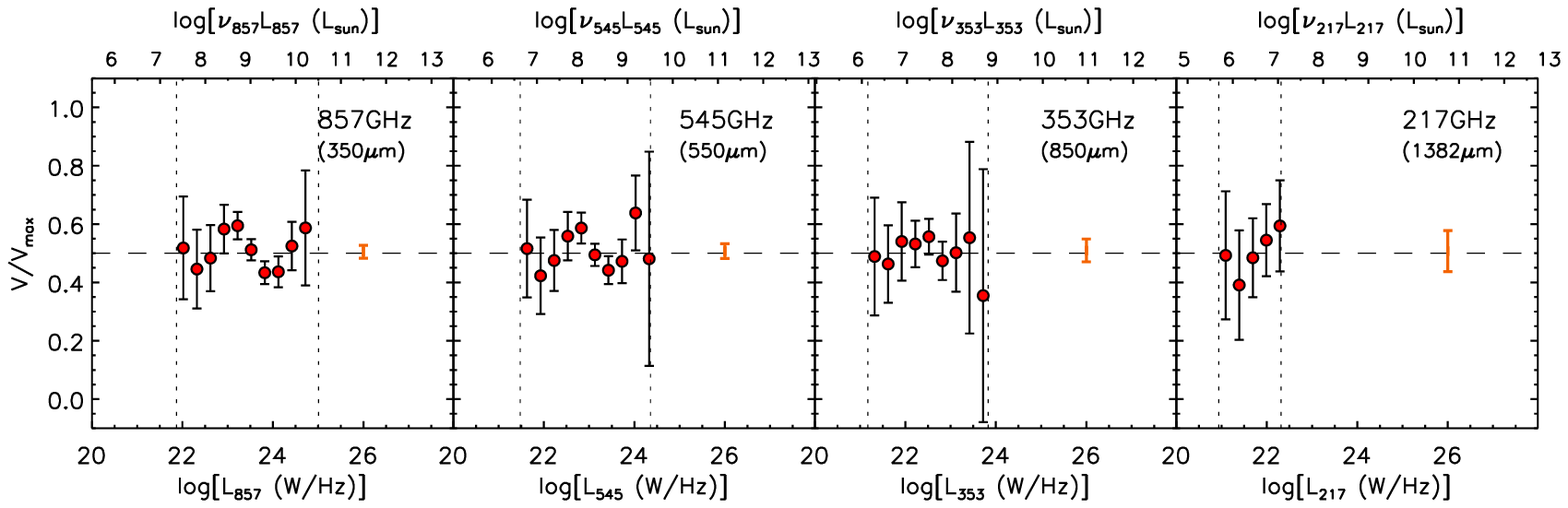}
}
\vspace{-5.7cm}
\caption{$V/V_{\rm max}$ test at 857\,GHz (350$\,\mu$m), 545\,GHz (550$\,\mu$m), 353\,GHz (850$\,\mu$m) and 217\,GHz (1382$\,\mu$m) for the flux-limited samples used to measure the luminosity function. The red dots, with error bars, are the mean $V/V_{\rm max}$ values for various luminosity bins while the orange error bar, here arbitrarily placed at a luminosity of $10^{26}\,$W/Hz, represents the $\pm 1\sigma$ uncertainty on the $V/V_{\rm max}$ value for the whole sample.  The vertical dotted lines have the same meaning as in Fig.\,\ref{fig:lf_Vmax}.}
 \label{fig:VonVmax}
\end{figure*}
%

\section{Luminosity functions}\label{sec:LF}

We have determined the luminosity functions using two different methods: the $1/V_{\rm  max}$ method and the parametric maximum likelihood method. Below we provide a brief description of the two approaches. The luminosity function is denoted as $\phi(L)$ and is taken to represent the comoving number density of objects per unit {\it logarithmic} interval in luminosity, i.e.
\begin{equation}
\phi(L) = \frac{dN}{dV\,d{\log}L}.
\end{equation}

\subsection{The $1/V_{\rm max}$ method}\label{sec:LF_Vmax}

\begin{figure*}
\vspace{0.5cm}
\hspace{+0.0cm}
\makebox[\textwidth][c]{
\includegraphics[width=0.95\textwidth]{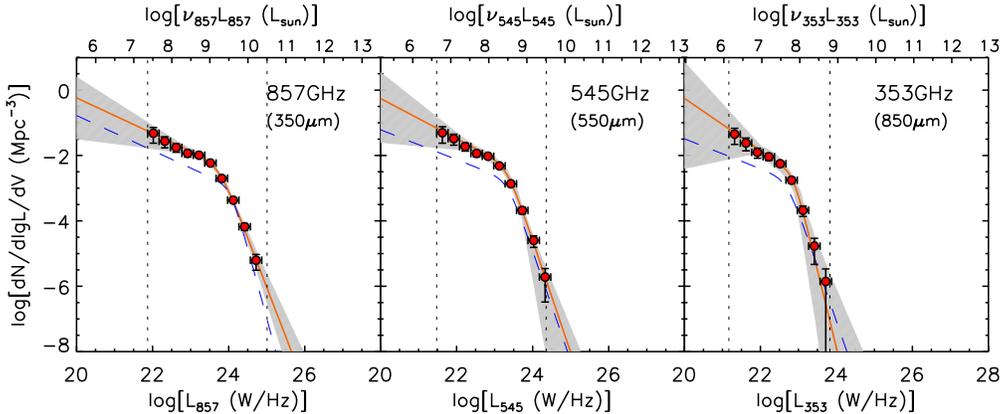}
}
\vspace{-5.7cm}
\caption{Local luminosity functions at 857\,GHz (350$\,\mu$m), 545\,GHz (550$\,\mu$m) and 353\,GHz (850$\,\mu$m) derived using the maximum likelihood parametric  method (orange curve) compared with that measured via the $1/V_{\rm max}$ method (red dots with error bars). The grey shaded regions correspond to the 68 per cent confidence interval for the parametric estimate. The predictions by Serjeant \& Harrison (2005) are also shown for comparison (dashed curve). The meaning of the vertical dotted lines is the same as in Fig.\,\ref{fig:lf_Vmax}.}
 \label{fig:lf_param}
\end{figure*}

The standard way of deriving the luminosity function is via the $1/V_{\rm max}$ estimator  (Schmidt 1968; Avni \& Bahcall 1980)
\begin{equation}\label{eq:LF_Vmax}
 \phi_{j} = \phi(L_{j}) =
  \frac{1}{\Omega ~\Delta{\log}L} \sum_{i}^{N_{j}}\frac{w_{i}}{V_{{\rm max}, i}},
\end{equation}
where $\Omega$ is the solid angle of the survey, $w_{i}$ are the weight factors taking into account the incompleteness of the samples (see \S\,\ref{subsec:diff_counts}) and the sum is over all the $N_{j}$ sources with luminosity in the range $[{\log}L_{j}-\Delta {\log}L/2$, ${\log}L_{j}+\Delta{\log}L/2]$. The quantity $V_{{\rm max}, i}$ represents the (comoving) volume, per unit solid angle, enclosed by the maximum (comoving) distance, $r_{{\rm max},i}$, at which the $i$-th object is detectable, given the survey flux limit, $F_{\rm lim}$, i.e., for the Euclidean geometry that applies here
\begin{equation}
V_{{\rm max}, i} = \frac{1}{3}r_{{\rm max},i}^{3},
\end{equation}
where
\begin{equation}
r_{{\rm max}, i} = \frac{1}{1+z_{i}} \left[ \frac{(1+z_{i})k(z_{i})L_{i} }{4\pi F_{\rm lim}}  \right]^{1/2}.
\end{equation}
As stated before, when no redshift-independent distance measurements are available we used the distance computed from the redshift of the source, i.e. $d_{i}=cz_{i}/H_{0}$, where $z_{i}$ has been corrected to the reference frame defined by the Cosmic Microwave Background radiation. However, as shown in Fig.~\ref{fig:Nz_Lrf_dist}, above the adopted completeness limit almost all the sources have an estimate of the distance that is independent of redshift.
In order to account for uncertainties in both flux densities and distances we adopted a bootstrap resampling method to derive the luminosity function. We have generated 1000 simulated catalogues by resampling, with repetitions, the input catalogue. In each simulation, a value of the flux density and of the distance is randomly assigned to each galaxy by assuming a Gaussian distribution with a mean equal to the measured values and $\sigma$ equal to the quoted errors. The rest-frame luminosities, the completeness correction factors and the maximum volumes are re-estimated from the simulated fluxes and distances. The luminosity function is then derived using eq.\,(\ref{eq:LF_Vmax}) considering only those objects with simulated flux density brighter than the adopted flux limit. Finally, the mean value of the simulated luminosity functions in each bin is taken as the estimate of the luminosity function in that bin and the rms of the distribution is taken as the uncertainty.
The luminosity function estimated with the $1/V_{\rm max}$ method is shown in Fig.\,\ref{fig:lf_Vmax} and tabulated in Table\,\ref{tab:LF_tabulated}. In a flux limited survey the faintest luminosities are undersampled and, as a consequence, the luminosity function derived from the bootstrap method exhibits an artificial turn-off at faint luminosities. Therefore we have adopted, in our analysis, a minimum luminosity, $L_{\rm min}$, corresponding to that of a source located 4\,Mpc away from us and
whose flux density is equal to the adopted flux density limit. This choice avoids the appearance of the turnoff. For similar reasons we excluded from the estimated luminosity function simulated luminosities higher than the maximum observed luminosity in the flux limited sample.
The adopted minimum and maximum luminosities are indicated by the vertical
dotted lines in Figs.\,\ref{fig:lf_Vmax}--\ref{fig:lf_param}.

The $1/V_{\rm max}$ estimator assumes a uniform spatial distribution of sources. In general this is not the case, and the effect of large scale structures may not average away, particularly when the sampled volume is relatively small.
In order to check if the distribution of the sources within each luminosity bin is consistent with uniformity, a $V/V_{\rm max}$ test was performed. In practice for each object within a given luminosity bin we computed the
ratio between the comoving volume up to the source position, $V_{i}$, and the corresponding maximum accessible volume $V_{{\rm max},i}$. For a uniform distribution $\langle V/V_{\rm max}\rangle = 0.5 \pm 1/\sqrt{12\,N}$, where $N$ is the number of sources in the bin. In order to account for the uncertainties in both flux density and distance measurements, the  $V/V_{\rm max}$ values have been computed for each of the simulated samples used to estimate the luminosity function and then averaged for each luminosity bin. The results are shown in Fig.\,\ref{fig:VonVmax}. In each panel, the orange error bar, arbitrarily placed at a luminosity of 10$^{26}\,$W/Hz, represents the $\pm 1\sigma$ uncertainty on the $\langle V/V_{\rm max}\rangle$ value for the whole sample. The latter is consistent with 0.5 for all the four samples, although some small deviations from 0.5 are observed in some luminosity bins at 857\,GHz. We conclude that, within 2$\,\sigma$, the distribution of galaxies in our samples is consistent with being statistically uniform. In fact, as {\it Planck} covers most of the sky, local inhomogeneities are significantly diluted.

%

\begin{table*}
\centering
\begin{tabular}{ccccccc}
\hline\hline
Band  & Band          &  $\log [N^{\prime} ({\rm Mpc}^{-3})]$ &  $\log[L_{\star} ({\rm W/Hz})]$  &  $\alpha$  & $\beta$ \\
(GHz) & ($\mu$m)  &  &   &   &   & \\
\hline
857 & 350    &  $-$2.28$_{-0.20}^{+0.16}$  &  23.76$_{-0.11}^{+0.13}$  & 0.54$_{-0.17}^{+0.13}$  &  3.03$_{-0.23}^{+0.33}$  \\
545 & 550    & $-$2.30$_{-0.50}^{+0.18}$  & 23.33$_{-0.12}^{+0.26}$  &  0.61$_{-0.17}^{+0.27}$  &  3.43$_{-0.31}^{+2.5}$ \\
353 & 850    & $-$2.54$_{-0.56}^{+0.38}$  &  22.88$_{-0.20}^{+0.25}$  &  0.80$_{-0.33}^{+0.38}$  &  4.76$_{-1.29}^{+2.40}$ \\
\hline
\hline
\end{tabular}
\caption{Parameters of the maximum likelihood luminosity functions [eq.~(\protect\ref{eq:LFparametric})] at 857\,GHz (350$\,\mu$m), 545\,GHz (550$\,\mu$m), 353\,GHz (850$\,\mu$m). At 217\,GHz (1382$\,\mu$m) the very limited statistic does not allow us to derive any meaningful constraint on the model parameters. The quoted errors correspond to the 68\% confidence intervals.}
\label{tab:lf_param}
\end{table*}
%

\subsection{Parametric Maximum Likelihood method}

A way to overcome the issue of large scale structure is to use maximum likelihood techniques, which can be either parametric or non-parametric (Sandage, Tammann \& Yahil 1979; Efstathiou, Ellis \& Peterson 1988; Saunders et al. 1990). The main assumption behind these methods is that the relative densities of galaxies of different luminosities are everywhere the same, and the normalization of the luminosity function is a local measure of density. In practice this means that the {\it observed} luminosity
function can be expressed as the product of a {\it universal} luminosity function, independent of position, and a position-dependent term that accounts for local inhomogeneities (Yahil, Sandage \& Tammann 1980; Yahil et al. 1991).

In this work we only implemented the parametric method (Sandage, Tammann \& Yahil 1979, STY method),  which has the advantage, compared with the non-parametric approach, that no binning of the data is required. However an analytic form of the luminosity function must be assumed. We adopt a double power-law functional form:
\begin{equation}\label{eq:LFparametric}
\phi(L|N^{\prime},L_{\star},\alpha,\beta) = N^{\prime} \left[ \left( \frac{L}{L_{\star}}\right)^{\alpha} + \left(  \frac{L}{L_{\star}} \right)^{\beta} \right]^{-1},
\end{equation}
as it provides a good fit to the local luminosity function of IRAS-selected galaxies \citep{SH05}. The double power-law has 4 free parameters in total: the normalization factor, $N^{\prime}$, the faint end slope, $\alpha$, the bright end slope, $\beta$, and the characteristic luminosity, $L_{\star}$, at which the transition from the two slopes occurs.
For a given set of values of the free parameters, the probability that a source is observed with rest-frame luminosity $L_{i}$, and with an {\it effective} redshift $z_{\rm eff,i}$ (i.e. the redshift that the source would have at the distance $d_{i}$ if peculiar motions were negligible) is thus given by
\begin{eqnarray}\label{eq:pi_parametric}
 p_{i} = \frac{1}{\tilde{N}} \, \phi(L_{i}|N^{\prime},L_{\star},\alpha,\beta) \, \frac{dV_{c}}{dzd\Omega}(z_{\rm
   eff,i}) \,\Omega
\end{eqnarray}
where $dV_{c}$ is the comoving volume element and $L_{{\rm min},i}$ is the minimum luminosity that can be observed
at the distance and the redshift of the $i-$th source, given the flux limit of the sample, i.e.
\begin{equation}
L_{{\rm min},i} = \frac{4\pi d_{i}^{2}F_{\rm lim}}{(1+z_{i})k(z_{i})}.
\end{equation}
$\tilde{N}$ is the total number of objects with $L\ge L_{{\rm min},i}$ and $F\ge F_{\rm lim}$ given the model LF
\begin{eqnarray}\label{eq:Ntilde}
\tilde{N} = \Omega \,\int_{z_{\rm min}}^{z_{\rm max}} dz
\frac{dV_{c}}{dzd\Omega}(z) \int_{  \mathpzc{L}_{\rm min}  }^{\infty}d{\log}L
\phi(L|N^{\prime},L_{\star},\alpha,\beta),
\end{eqnarray}
where $\mathpzc{L}_{\rm min}=\max[{\log}L_{\rm min,i},{\log}L_{\rm lim}(z)]$, and $L_{\rm lim}(z)$ is the luminosity corresponding to the flux density limit at redshift $z$, while $z_{\rm min}$ and $z_{\rm max}$ are the minimum and the maximum redshifts observed in the flux-/luminosity-limited sample. \\
The probability defined by eq.\,(\ref{eq:pi_parametric}) is independent of the normalization $N^{\prime}$ as the latter cancels out in the ratio between the differential and cumulative luminosity functions. This is what makes the maximum likelihood approach independent of local inhomogeneities under the assumption of a {\it universal} luminosity function. The probability of observing $N$ objects with luminosities $\left\{ L_{1},L_{2}, ...,L_{N} \right\}$, with redshifts $\left\{z_{1}, z_{2}, ..., z_{N} \right\}$ and with physical distances $\left\{ d_{1}, d_{2}, ..., d_{N} \right\}$ is
\begin{eqnarray}
P = \frac{\tilde{N}^{N}e^{-\tilde{N}}}{N!}\, \prod_{i=1}^{N}p_{i}
\end{eqnarray}
where we have assumed that the number of detected sources follows a Poisson distribution with expectation value equal to $\tilde{N}$ defined by eq.\,(\ref{eq:Ntilde}). In order to account for the incompleteness in our samples we follow Patel et al. (2012), by replacing the total number of objects ($N$) with $\sum w_{i}$ and introducing a weighting term, $w_{i}/\langle w \rangle$ ($\langle w\rangle$ being the average weight) as an exponent of the individual source likelihoods, such that
\begin{equation}\label{eq:completeness}
P \propto \tilde{N}^{^{\sum{w_{i}}}}e^{-\tilde{N}}\,\prod_{i=1}^{N}p_{i}^{\frac{w_{i}}{\langle w\rangle}}.
\end{equation}
By maximizing the probability $P$ (i.e. the likelihood of the sample) one obtains the set of values of $N^{\prime}$, $L_{\star}$, $\alpha$ and $\beta$ that make the observed data set the ``most probable'' one. Following the same bootstrap resampling method used to measure the luminosity function via the $1/V_{\rm max}$ method, we have estimated the maximum likelihood values of the free
parameters for each of the simulated samples.
The derived luminosity functions and their uncertainties are shown in Fig.\,\ref{fig:lf_param}. They are fully consistent with those derived from the $1/V_{\rm max}$ method. We have checked that the consistency with the $1/V_{\rm max}$ results is mantained even if we use a power-law/Gaussian analytic model\footnote{In this model the LF behaves as a power law for $L<<L_{\star}$ and as a Gaussian in $\log L$ for $L>>L_{\star}$} for the LF, like the one suggested by Saunders et al. (1990),
\begin{eqnarray}
\phi(L|N^{\prime},L_{\star},\alpha,\sigma) = N^{\prime}\left( \frac{L}{L_{\star}} \right) \exp\left[ -\frac{1}{2\sigma^2}\log^{2}\left( 1 +\frac{L}{L_{\star}}\right) \right]. \nonumber \\
~~
\end{eqnarray}
The agreement between the two estimates further supports the conclusion that inhomogeneities associated with large scale structure in the local Universe are not an issue in our analysis. Besides, overdensities associated with galaxy clusters in the local Universe are less pronounced in far-infrared/sub-mm surveys than in optical/near-infrared surveys. In fact such regions, especially their dense core, are dominated by ellipticals which have relatively low dust emission. The maximum likelihood values of the parameters and their 68\% confidence intervals, derived from the  distribution of values produced by the bootstrap, are given in Table\,\ref{tab:lf_param} for the 3 higher frequencies (857, 545 and 353 GHz). As illustrated by Fig.\,\ref{fig:lf_param} the luminosity range covered by the 217 GHz sample is too small to allow a reliable estimate of the slopes $\alpha$ and $\beta$. 

As a sanity check, we use eq.\,(6) of Planck Collaboration VII (2012) to derive the normalization of the Euclidean plateau of the number counts from the estimated luminosity function  and compare it with the values quoted in \S\,\ref{subsec:diff_counts}. The results, based on the best-fit parameters listed in Table\,\ref{tab:lf_param}, are 509, 109 and 14\,Jy$^{1.5}$\,sr$^{-1}$ at 857, 545 and 353\,GHz, respectively, in excellent agreement with those derived from the observed number counts.

\begin{figure}
\vspace{+0.5cm}
\hspace{-7.0cm}
\makebox[\textwidth][c]{
\includegraphics[width=0.8\textwidth]{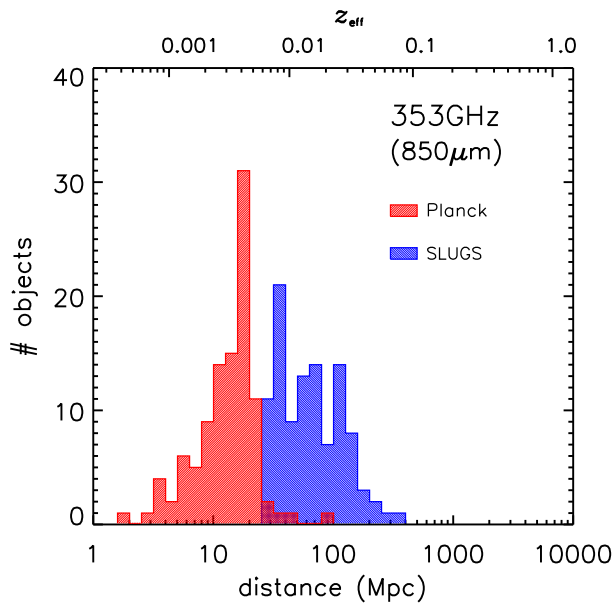}
}
\vspace{-1.4cm}
\caption{Distribution of source (physical) distances in the SLUGS sample (blue histogram) and in the {\it Planck} flux-limited sample used to measure the luminosity function (red histogram). The  value of the redshift corresponding to a given distance ({\it effective} redshift, see text) is shown in the upper x-axis.}
 \label{fig:Planck_vs_SLUGS_dist}
\end{figure}

\section{Comparison with earlier estimates}\label{sec:results}

\subsection{Comparison with Dunne et al. (2000) and Serjeant \& Harrison (2005)}

Dunne et al. (2000) have carried out a follow-up program at 850\,$\mu$m with SCUBA $-$ the SCUBA Local Universe Galaxy Survey (SLUGS hereafter) $-$ of a sample of galaxies selected from the {\it IRAS} Bright Galaxy Sample  (BGS; Soifer et al. 1989). The BGS is complete to a 60$\,\mu$m flux limit of 5.24\,Jy at $|b|>30^{\circ}$ and $\delta>-30^{\circ}$. The  SLUGS sample includes all BGS galaxies with declination in the range $-10^{\circ}<\delta<50^{\circ}$ and with velocities above $v_{\rm min}=1900\,$km\,s$^{-1}$ (corresponding
to a minimum redshift $z_{\rm min}=v_{\rm min}/c=0.0063$). The minimum velocity is chosen to exclude those galaxies whose angular size exceeds the SCUBA field-of-view ($\sim$2 arcminutes). In Fig.\,\ref{fig:Planck_vs_SLUGS_dist} we compare the distribution of source distances in the SLUGS sample with that of the {\it Planck} flux-/luminosity-limited sample from
which the luminosity function has been measured. The peak observed around 17\,Mpc is partially due to the Virgo cluster. However we have checked that the estimated luminosity functions do not change significantly if the region associated with the Virgo Cluster is masked off. 

\begin{figure}
\vspace{+0.0cm}
\hspace{-6.5cm}
\makebox[\textwidth][c]{
\includegraphics[width=0.78\textwidth]{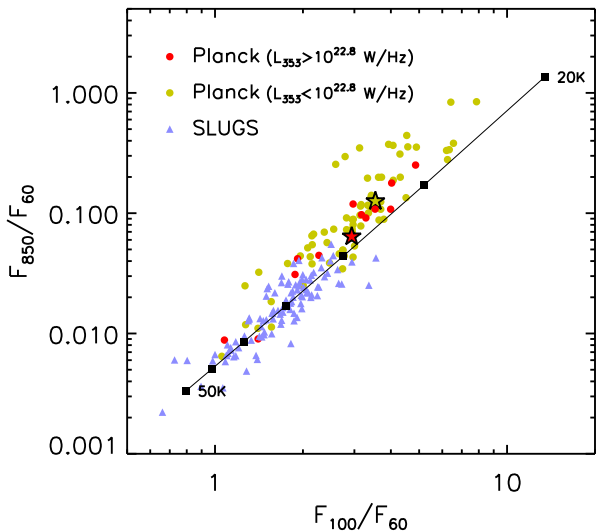}
}
\vspace{-1.4cm}
\caption{Sub-mm/far-infrared colour-colour plane for the 353\,GHz {\bf (850$\,\mu$m)} selected {\it Planck} sources with completeness above 80\% and with luminosity $<\,$10$^{22.8}\,$W/Hz (red dots) or $>\,$10$^{22.8}\,$W/Hz (yellow dots). The stars represent the mean values. The SLUGS sample (Dunne et al. 2000; blue triangles) is shown for comparison. The solid line is the track for a grey-body with dust emissivity index $\beta=1.3$ and temperatures ranging from 20 up to 50\,K in steps of 0.5\,K. }
 \label{fig:S850onS60_vs_S100onS60}
\end{figure}

The overlap between the {\it Planck} and the SLUGS samples is poor (only 5 galaxies are in common) mainly because of the low-$z$ cut adopted by Dunne et al. (2000). The luminosity function derived by the latter authors is shown by the blue squares in Fig.\,\ref{fig:lf_Vmax}. It agrees with the {\it Planck} measurements for $L_{353\rm GHz}\gsim10^{23}\,$W/Hz but lies significantly below them at fainter luminosities. A likely explanation of the difference is the bias against cold dusty galaxies implicit in the SLUGS sample. In fact, the selection at 60\,$\mu$m combined with the imposed minimum redshift tends to favor galaxies with relatively bright infrared luminosities which usually have warmer SEDs (e.g. Smith et al. 2011). This bias has been demonstrated by Vlahakis et al. (2005) and, more recently, by Planck Collaboration XVI (2011) who have performed a detailed analysis of the SED properties of a sample of low-redshift galaxies extracted from the ERCSC. We confirm those results by comparing the sub-mm/far-infrared colours of the SLUGS sample with those of the {\it Planck} sources used to derive the luminosity function at 850\,$\mu$m. The results are shown in Fig.\,\ref{fig:S850onS60_vs_S100onS60} together with the track of a grey-body with dust emissivity index $\beta=1.3$ and temperatures ranging from 20 to 50\,K (in steps of 5\,K). The SLUGS sample has higher dust temperatures than the 850$\,\mu$m {\it Planck}-selected sample, and the difference is more pronounced for galaxies with lower 850\,$\mu$m luminosities (yellow dots). 

Serjeant \& Harrison (2005) have gone one step further and derived the local luminosity function of galaxies at many wavelengths from $70\,\mu$m to $850\,\mu$m by modeling the SEDs of all 15411 galaxies from the redshift survey of the IRAS Point Source Catalogue (PSC$z$; Saunders et al. 2000). The PSC$z$ survey covers 84 per cent of the sky to a depth of 0.6\,Jy at 60$\,\mu$m and has a median redshift of 8400\,km s$^{-1}$ (i.e. $z=0.028$). Starting from the {\it IRAS} measurements at 60 and 100$\,\mu$m, Serjeant \& Harrison have exploited the strong correlation observed in the SLUGS between the far-infrared luminosity and the sub-mm/far-infrared colour to predict the sub-mm fluxes of each of the PSC$z$ galaxies. As stated by the authors, these predictions are sufficient to define the sub-mm fluxes to within a factor of 2. The Serjeant \& Harrison results at {\it Planck} wavelengths are shown by the dashed curves in Figs.~\ref{fig:dNdlgF_euc} and \ref{fig:NgtF} for the number counts and in Figs.~\ref{fig:lf_Vmax} and \ref{fig:lf_param} for the luminosity functions. Not surprisingly, since again we are dealing with an IRAS-selected sample, the agreement is good at high luminosities but {\it Planck}-based estimates are higher below the characteristic luminosity $L_{\star}$.

\begin{figure}
\vspace{+0.0cm}
\hspace{-6.5cm}
\makebox[\textwidth][c]{
\includegraphics[width=0.78\textwidth]{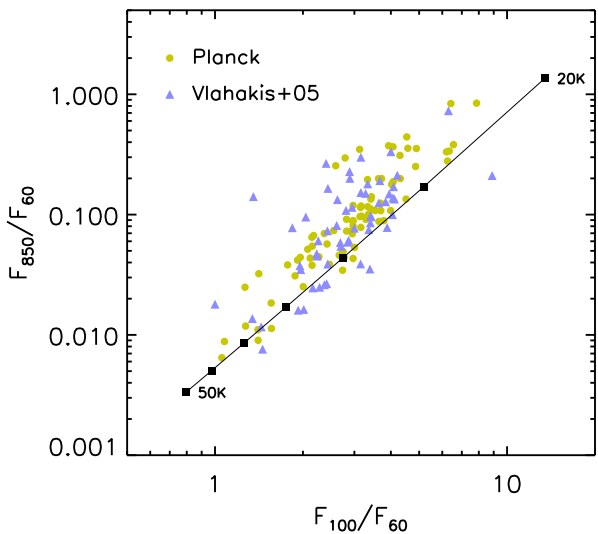}
}
\vspace{-1.4cm}
\caption{Sub-mm/far-infrared colour-colour plane for the 353\,GHz {\bf (850$\,\mu$m)} selected {\it Planck} sources with completeness above 80\% (yellow dots) and for the sample of optically selected galaxie of Vlahakis et al. (2005; blue triangles). The solid line is the the same as in Fig.\,\ref{fig:S850onS60_vs_S100onS60}.}
 \label{fig:S850onS60_vs_S100onS60_Vlahakis}
\end{figure}

\subsection{Comparison with Vlahakis et al. (2005)}

To check whether the IRAS selection misses populations of sub-millimeter emitting galaxies Vlahakis et al. (2005) observed with SCUBA a sample of 81 galaxies selected from the Center for Astrophysics (CfA) optical redshift survey (Huchra et al. 1983). They found that the ratios of the mass of cold dust to the mass of warm dust for these galaxies is much higher than for the SLUGS galaxies, thus demonstrating that the SLUGS was missing a significant population of galaxies characterized by large proportions of cold dust. In fact, the 850$\,\mu$m luminosity function derived by Vlahakis et al. is higher than the one derived from the SLUGS.

The green dots in the 850$\,\mu$m panel of Fig.\,\ref{fig:lf_Vmax} show the luminosity function of Vlahakis et al. obtained with the method of Serjeant \& Harrison, but using sub-mm/far-IR colour relations re-calibrated on the SLUGS {\it and} the optically selected samples together\footnote{The 850$\,\mu$m luminosity function derived by Vlahakis et al.  using the method of Serjeant \& Harrison is in perfect agreement with that directly measured from the optically selected sample alone. We decided to show the former because it has a better statistic.}. The result is in better agreement with the ERCSC luminosity function, and, as expected,
the sub-mm/far-infrared colours of the Vlahakis et al. sample are consistent with those of the 850$\,\mu$m {\it Planck}-selected sample, as shown in Fig.\,\ref{fig:S850onS60_vs_S100onS60_Vlahakis}.

\subsection{Comparison with Vaccari et al. (2010)}

Vaccari et al. (2010) have estimated the local luminosity functions of galaxies selected at 250, 350 and 500\,$\mu$m in the Lockman Hole and XFLS fields (14.7\,deg$^{2}$) observed during the Herschel Science Demonstration Phase (SDP) of  HerMES (the {\it Herschel} Multi-tiered Extragalactic Survey; Oliver et al. 2012).
Their samples cover the redshift range $0<z<0.2$ and comprise 210 sources at $350\,\mu$m and 45 sources at $500\,\mu$m; at both wavelengths they have spectroscopic redshifts for $\simeq 85\%$ of their sources.

We have converted to 545 GHz ($550\,\mu$m) their $500\,\mu$m luminosity function by assuming $\alpha=2.7$, as this is the mean value of the sub-mm spectral index found from the analysis of the SED of local galaxies in the {\it Planck} sample (Clemens et al., in prep.).  Their results, shown by the blue triangles in Fig.\,\ref{fig:lf_Vmax}, generally agree, within the uncertainties, with ours, even though Vaccari et al. did not take into account any evolutionary correction, while significant evolution between  $0 <z <0.1$ and $0.1 <z <0.2$ was reported, at $250\,\mu$m, by Dye et al. (2010) and Dunne et al. (2011).

The estimates of the local (sub-)mm luminosity function presented here are the first derived from a complete all-sky (sub-)mm selected sample of galaxies. Together with the estimates of the luminosity functions of higher redshifts galaxies produced by {\it Herschel} (Eales et al. 2010b; Gruppioni et al. 2010; Lapi et al. 2011), they will provide a critical constraint for any model of galaxy evolution.

\section{Conclusions}\label{sec:conclusions}

The {\it Planck} ERCSC has provided the first samples of truly local ($\hbox{distance}\lsim 100\,$Mpc) galaxies blindly selected at (sub)-mm wavelengths, large enough to allow us to obtain accurate determinations of the local luminosity function at 857, 545 and 353\,GHz down to luminosities one order of magnitude fainter than previous estimates and well below the characteristic luminosity $L_\star$. We have also obtained the first estimate, albeit over a limited luminosity range, of the local luminosity function of dusty galaxies at 217\,GHz. To get there several steps were necessary:
\begin{itemize}
\item Identification of the ERCSC flux density measurement most appropriate for our galaxies and application of suitable corrections,

\item separation of dusty sources from radio loud AGNs,

\item removal of Galactic sources,

\item determination of the completeness limits at each frequency,

\item correction of {\it Planck} fluxes for the contribution of $CO$ lines.

\end{itemize}
Although spectroscopic redshifts are available for all galaxies in our samples, they may not be good distance indicators because of the potentially important contributions of peculiar motions. Therefore we have used, as far as possible, redshift-independent distance estimates that are available for most sources. The effect of local inhomogeneities in the galaxy distribution was investigated by means of the $V/V_{\rm max}$ test in different luminosity bins. We also used, in addition to the classical $1/V_{\rm max}$ estimator, a parametric maximum likelihood estimator of the luminosity function, that is immune to the effect of inhomogeneities.

The derived luminosity functions are in good agreement, at high luminosities, with those obtained using follow-up observations of $60\,\mu$m selected IRAS galaxies (Serjeant \& Harrison 2005; Dunne et al. 2000). Below the characteristic luminosity, $L_\star$, however, our estimates are significantly higher, due to the bias against cold, generally low-luminosity galaxies, inherent in the SLUGS sample, and close the the estimate by Vlahakis et al. (2005) based on SCUBA follow-up of an optically selected sample. Our estimates are also consistent with those by Vaccari et al. (2010) based on data from the HerMES survey at 350 and $500\,\mu$m.

A detailed analysis of the astrophysical properties (such as star-formation rates, dust masses, dust temperatures, infrared luminosity function etc.) of the galaxies used here to derive the local sub-mm luminosity function  will be presented in a forthcoming paper (Clemens et al. in prep.).

\section*{Acknowledgements}
This work was supported by ASI/INAF agreement I/072/09/0 and by PRIN MIUR 2009. We thank the anonymous referee for useful comments.
J.G.N. acknowledges financial support from Spanish CSIC for a JAE-DOC fellowship. L.T.,  J.G.N. and L.B. acknowledge partial financial support from the Spanish Ministerio de Ciencia e Innovacion under project AYA2010-21766-C03-01. This research has made use of Aladin and of the NASA/IPAC Extragalactic Database (NED) which is operated by the Jet Propulsion Laboratory, California Institute of Technology, under contract with the National Aeronautics and Space Administration.


\begin{thebibliography}{}


\bibitem[\protect\citeauthoryear{Avni \& Bahcall}
{1980}]{AB80} Avni Y. \& Bahcall J.~N., 1980, ApJ, 235, 694


\bibitem[\protect\citeauthoryear{Beichman et al.}{1988}]{Beichman88} Beichman C.~A., Neugebauer G., Habing H.~J., Clegg P.~E., Chester T.~J., 1988, IRAS Explanatory Supplement



\bibitem[\protect\citeauthoryear{Bonnarel et al.}{2000}]{Aladin00} Bonnarel F. et al., 2000, 143, 33

\bibitem[\protect\citeauthoryear{Boselli et al.}{2010}]{Boselli10} Boselli A. et al., 2010, PASP, 122, 261

\bibitem[\protect\citeauthoryear{Ciesla et al.}{2012}]{Ciesla12} Ciesla L. et al., 2012, A\&A, in press (astro-ph/1204.4726)

\bibitem[\protect\citeauthoryear{Clements et al.}{2010}]{Clements10} Clements D.~L., Dunne L. \& Eales S., 2010, MNRAS, 403, 274

\bibitem[\protect\citeauthoryear{Daddi et al.}{2010}]{Daddi10} Daddi E. et al., 2010, ApJL, 714, 118

\bibitem[\protect\citeauthoryear{Dale et al.}{2012}]{Dale12} Dale D.~A., et al., 2012, ApJ, 745, 95

\bibitem[\protect\citeauthoryear{Davies et al.}{2012}]{Davies12} Davies J.~I., et al., 2012, MNRAS, 419, 3505


\bibitem[\protect\citeauthoryear{Dunne et al.}{2011}]{Dunne2011} Dunne L., et al., 2011, MNRAS, 417, 1510

\bibitem[\protect\citeauthoryear{Dunne et al.} {2000}]{Dunne00} Dunne L., Eales S., Edmunds M., Ivison R., Alexander P., Clements D.~L., 2000, MNRAS, 315, 115

\bibitem[\protect\citeauthoryear{Dye et al.}{2010}]{Dye10} Dye S. et al., 2010, A\&A, 518, 10





\bibitem[\protect\citeauthoryear{Eales et al.} {2010a}]{Eales10a} Eales S. et al., 2010, PASP, 122, 499

\bibitem[\protect\citeauthoryear{Eales et al.}{2010b}]{Eales10b} Eales S. et al. 2010, A\&A, 518, L23

\bibitem[\protect\citeauthoryear{Eddington}{1913}]{1913MNRAS..73..359E} Eddington A.~S., 1913, MNRAS, 73, 359

\bibitem[\protect\citeauthoryear{Efstathiou, Ellis, \& Peterson}{1988}]{Efstathiou88} Efstathiou G., Ellis R.~S., Peterson B.~A., 1988, MNRAS, 232, 431



\bibitem[\protect\citeauthoryear{Gehrels} {1986}]{G86} Gehrels N., 1986, ApJ, 303, 336

\bibitem[\protect\citeauthoryear{Genzel et al.}{2010}]{Genzel10} Genzel R. et al. 2010, MNRAS, 407, 2091

\bibitem[\protect\citeauthoryear{Gruppioni et al.}{2010}]{Gruppioni10}  Gruppioni C. et al., 2010,A\&A, 518, 27

\bibitem[\protect\citeauthoryear{Gruppioni et al.}{2011}]{Gruppioni11}  Gruppioni C., Pozzi  F., Zamorani, G., \& Vignali C. 2011, MNRAS, 416, 70





\bibitem[\protect\citeauthoryear{Healey et al.}{2007}]{Healey07} Healey S.~E., Romani R.~W., Taylor G.~B., Sadler E.~M., Ricci R., Murphy T., Ulvestad J.~S., Winn J.~N., 2007, ApJS, 171, 61

\bibitem[\protect\citeauthoryear{Huchra et al.}{1983}]{Huchra83} Huchra J., Davis M., Latham D., Tonry J., 1983, ApJS, 52, 89

\bibitem[\protect\citeauthoryear{Herranz et al.}{2012}]{Herranz2012} Herranz D., et al., 2012, A\&A submitted, arXiv:1204.3917



\bibitem[\protect\citeauthoryear{Lapi et al.}{2011}]{Lapi11} Lapi A. et al., 2011, ApJ, 742, 24

\bibitem[\protect\citeauthoryear{Leroy et al.}{2009}]{Leroy09}  Leroy A.~K. et al., 2009, AJ, 137, 4670

\bibitem[\protect\citeauthoryear{Mao et al.}{2010}]{Mao2010} Mao R.-Q., Schulz A., Henkel C., Mauersberger R., Muders D., Dinh-V-Trung, 2010, ApJ, 724, 1336

\bibitem[\protect\citeauthoryear{Moshir et al.} {1992}]{Moshir92} Moshir M. et al., 1992, Explanatory Supplement to the IRAS Faint Source Survey, Version 2, JPL D-10015 8/92 JPL, Pasadena



\bibitem[\protect\citeauthoryear{Negrello et al.}{2004}]{Neg04} Negrello M. et al., 2004, MNRAS, 352, 493

\bibitem[\protect\citeauthoryear{Negrello et al.}{2005}]{Neg05} Negrello M. et al., 2005, MNRAS, 358, 869



\bibitem[\protect\citeauthoryear{Oliver et al.}{2012}]{Oliver2012} Oliver S. et al., 2012, arXiv:1203.2562



\bibitem[\protect\citeauthoryear{Papadopoulos et al.}{2011}]{Papadopoulos2011} Papadopoulos P.~P., van der Werf P., Xilouris E.~M., Isaak K.~G., Gao Y., Muehle S., 2011, arXiv:1109.4176

\bibitem[\protect\citeauthoryear{Patel et al.} {2012}]{Patel12} Patel H., Clements D.~L., Vaccari M., Mortlock D.~J., Rowan-Robinson M. \& Perez-Fournon I., 2012, MNRAS in press (astro-ph/1205.5690)




\bibitem[\protect\citeauthoryear{Pilbratt et al.}{2010}]{Pil10} Pilbratt G. et al., 2010, A$\&$A, 518, 1


\bibitem[\protect\citeauthoryear{Planck Collaboration}{2011}]{ExplanSuppl} Planck Collaboration, 2011, The Explanatory Supplement to the Planck Early Release Compact Source Catalogue (ESA)

\bibitem[\protect\citeauthoryear{Planck Collaboration I}{2011}]{PIP_I} Planck Collaboration I, 2011, A\&A, 536, A1

\bibitem[\protect\citeauthoryear{Planck Collaboration VII}{2011}]{PIP_VII} Planck Collaboration VII, 2011, A\&A, 536, A7

\bibitem[\protect\citeauthoryear{Planck Collaboration XIII}{2011}]{PIP_XIII} Planck Collaboration XIII, 2011, A\&A, 536, A13

\bibitem[\protect\citeauthoryear{Planck Collaboration XIV}{2011}]{ERCSCvalidation} Planck Collaboration XIV, 2011, A\&A, 536, A14

\bibitem[\protect\citeauthoryear{Planck Collaboration XVI}{2011}]{PIP_XVI} Planck Collaboration XVI, 2011, A\&A, 536, A16

\bibitem[\protect\citeauthoryear{Planck Collaboration VII}{2012}]{PIP_XXXVIII} Planck Collaboration VII, 2012, A\&A submitted, arXiv:1207.4706


\bibitem[\protect\citeauthoryear{Polletta et al.}{2007}]{Polletta2007} Polletta M., et al., 2007, ApJ, 663, 81



\bibitem[\protect\citeauthoryear{Sandage, Tammann, \& Yahil}{1979}]{Sandage79} Sandage A., Tammann G.~A., Yahil A., 1979, ApJ, 232, 352

\bibitem[\protect\citeauthoryear{Saunders et al.}{1990}]{Saunders90} Saunders W., Rowan-Robinson M., Lawrence A., Efstathiou G., Kaiser N., Ellis R.~S., Frenk C.~S., 1990, MNRAS,
242, 318

\bibitem[\protect\citeauthoryear{Saunders et al.}{2000}]{Saunders00} Saunders W. et al., 2000, MNRAS, 317, 55

\bibitem[\protect\citeauthoryear{Sanders et al.}{2000}]{Sanders03} Sanders D.~B., Mozzarella J.~M., Kim D.~C., Surace J.~A. \& Soifer B.~T., AJ, 126, 1607


\bibitem[\protect\citeauthoryear{Schlegel, Finkbeiner, \& Davis}{1998}]{Schlegel1998} Schlegel D.~J., Finkbeiner D.~P., Davis M., 1998, ApJ, 500, 525

\bibitem[\protect\citeauthoryear{Schmidt}{1968}]{Schmidt68} Schmidt M., 1968, ApJ, 151, 393


\bibitem[\protect\citeauthoryear{Serjeant \& Harrison}{2005}]{SH05} Serjeant S. \& Harrison D., 2005, MNRAS, 356, 192



\bibitem[\protect\citeauthoryear{Smith et al.}{2011}]{Smith2011} Smith D.~J.~B., et al., 2011, MNRAS, 416, 857


\bibitem[\protect\citeauthoryear{Soifer et al.}
{1989}]{Soifer89} Soifer B.~T., Boehmer L., Neugebauer G. \& Sanders
D.~B., 1989, AJ, 98, 766

\bibitem[\protect\citeauthoryear{Solomon et al.}{1997}]{Solomon1997} Solomon P.~M., Downes D., Radford
S.~J.~E., Barrett J.~W., 1997, ApJ, 478, 144



\bibitem[\protect\citeauthoryear{Vaccari et al.}
{2010}]{Vaccari10} Vaccari M. et al., 2010, A\&A, 518, 20

\bibitem[\protect\citeauthoryear{Vlahakis et al.}
{2005}]{Vlahakis05} Vlahakis C., Loretta D. \& Eales S., 2005, MNRAS, 364, 1253





\bibitem[\protect\citeauthoryear{Wang \& Rowan-Robinson}
{2009}]{Wang09} Wang L. \& Rowan-Robinson M., 2009, MNRAS, 398, 109

\bibitem[\protect\citeauthoryear{Wright et al.}{2010}]{Wright10} Wright E.~L., et al., 2010, AJ, 140, 1868

\bibitem[\protect\citeauthoryear{Yahil, Sandage, \& Tammann}{1980}]{Yahil80} Yahil A., Sandage A., Tammann G.~A., 1980, ApJ, 242, 448

\bibitem[\protect\citeauthoryear{Yahil et al.}{1991}]{Yahil91} Yahil A., Strauss M.~A., Davis M., Huchra J.~P., 1991, ApJ, 372, 380

\bibitem[\protect\citeauthoryear{Yao et al.} {2003}]{Yao03} Yao L., Seaquist E.~R., Kuno N., Dunne L., 2003, ApJ, 588, 771

\end{thebibliography}
\end{document}